\documentclass[12pt,a4]{article}
\usepackage{amsthm,amsmath,latexsym,amssymb,amsfonts,amscd}
\usepackage{graphics,fancyhdr,array,stmaryrd,euscript}
\usepackage{slashed,tikz}
\numberwithin{equation}{section}

\usepackage{hyperref}

\usepackage[numbers,sort&compress]{natbib}
\usepackage[nottoc]{tocbibind}
\usepackage{xcolor}

\usepackage[all,cmtip]{xy}

\newcommand{\TikzRect}[2]{\filldraw[color=black,fill=red]  (#1-\R,#2-\R) rectangle (#1+\R,#2+\R);}

\newcommand{\TikzRectG}[2]{\filldraw[color=black,fill=green]  (#1-\R,#2-\R) rectangle (#1+\R,#2+\R);}

\newcommand{\FIELDS}{
\begin{tikzpicture}[scale=0.7]
\tikzset{%
  >=latex, 
  inner sep=0pt,%
  outer sep=2pt,%
  mark coordinate/.style={inner sep=0pt,outer sep=0pt,minimum size=3pt,
    fill=black,circle}%
}
\def\R{0.15}
\def\mar{0.15}

\draw[->,black,thick] (0,0) -- (0,9) node[left]{$\# A$ };
\draw[->,black,thick] (0,0) -- (11,0) node[right]{$\# A'$} ;

\filldraw[color=black,fill=green]  (12,6) circle (\R);
\node[right] at (12.3,6) {one-forms, $\omega$};

\TikzRectG{12}{7}  \TikzRect{12.5}{7}  
\node[right] at (12.7,7) {zero-forms, $C$};

\filldraw[color=black,fill=green]  (4,0) circle (\R);
\filldraw[color=black,fill=green]  (3,1) circle (\R);
\filldraw[color=black,fill=green]  (2,2) circle (\R);
\filldraw[color=black,fill=green]  (1,3) circle (\R);
\filldraw[color=black,fill=green]  (0,4) circle (\R);
\draw[green,thick] (4,0) -- (0,4);

    \draw[-latex,thick] (4.5,5.5) node[right,scale=1.0]{$\omega^{A(2s-2)}$}
        to[out=-180,in=50] (+\mar, 4);
        
    \draw[-latex,thick] (4.5,8) node[right,scale=1.0]{$\Psi^{A(2s)}$, $\lambda=-s$ Weyl tensor}
        to[out=-120,in=10] (+\mar,5);

    \draw[-latex,thick] (8.5,1.5) node[right,scale=1.0]{$C^{A'(2s)}$, $\lambda=+s$ Weyl tensor}
        to[out=180,in=-45] (5, -\mar);
        
    \draw[rounded corners] ( -0.5 , 3.5 ) rectangle (0.5, 5.5) {};        

    \draw[-latex,thick] (2,-1) node[left,scale=1.0]{$\omega^{A'(2s-2)}$}
        to[out=0,in=-90] (4, -\mar);

    \TikzRect{0}{5}
    \TikzRect{1}{6}
    \TikzRect{2}{7}
    \TikzRect{3}{8}
    \draw[red,thick] (0,5) -- (4,9);

    \TikzRectG{5}{0}
    \TikzRectG{6}{1}
    \TikzRectG{7}{2}
    \TikzRectG{8}{3}
    \draw[green,thick] (5,0) -- (9,4);

    \filldraw[color=black,fill=black]  (4.5-1.5*\R,-\R) rectangle (4.5+1.5*\R,+\R);

    \draw[-latex,thick] (4.5,3.5) node[above,scale=1.0]{\large $\mathcal{V}(e,e,C)$-cocycle}
        to[out=-90,in=90] (4.5, 0.5-\mar);

\end{tikzpicture}}

\newcommand{\pl}{\partial}
\newcommand{\plb}{\bar{\partial}}
\newcommand{\be}{\begin{equation}}
\newcommand{\ee}{\end{equation}}
\newcommand{\bea}{\setlength\arraycolsep{2pt} \begin{eqnarray}}
\newcommand{\eea}{\end{eqnarray}}

\newcommand{\hs}{{\mathfrak{hs}}}

\newcommand{\ga}{A}

\newcommand{\gad}{{A'}}

\newcommand{\fud}[2]{{}^{#1}{}_{#2}\,}
\newcommand{\fdu}[2]{{}_{#1}{}^{#2}\,}

\newcommand{\fdud}[3]{{}_{#1}{}^{#2}{}_{#3}\,}

 \newcommand{\bry}{{{\bar{y}}}}

\newcommand{\besubeqs}{\begin{subequations}}
\newcommand{\esubeqs}{\end{subequations}}

\newcommand{\xx}{{\rho}}

\newcommand{\Tr}{\mathrm{Tr}}

\begin{document}
\pagenumbering{gobble}
\hfill
\vskip 0.01\textheight
\begin{center}
{\Large\bfseries 
Low spin solutions of Higher Spin Gravity: \\ [7pt] BPST instanton}

\vskip 0.03\textheight
\renewcommand{\thefootnote}{\fnsymbol{footnote}}
Evgeny \textsc{Skvortsov}\footnote{Research Associate of the Fund for Scientific Research -- FNRS, Belgium}\footnote{Also at Lebedev Institute of Physics}\footnote{\label{note1}First authors, in alphabetical order.}${}^{a}$ \& Yihao \textsc{Yin}\footref{note1}${}^{b,c}$
\renewcommand{\thefootnote}{\arabic{footnote}}
\vskip 0.03\textheight

{\em ${}^{a}$ Service de Physique de l'Univers, Champs et Gravitation, \\ Universit\'e de Mons, 20 place du Parc, 7000 Mons, 
Belgium}\\ \vspace*{5pt}
{\em ${}^{b}$ College of Physics, Nanjing University of Aeronautics and Astronautics, \\
Nanjing 211106, China}\\
{\em ${}^{c}$ Key Laboratory of Aerospace Information Materials and Physics (NUAA), MIIT,\\
Nanjing 211106, China}\\

\vskip 0.05\textheight

\begin{abstract}
Higher spin gravities do not have a low energy limit where higher-spin fields decouple from gravity. Nevertheless, it is possible to construct fine-tuned exact solutions that activate low-spin fields without sourcing the higher-spin fields. We show that BPST (Belavin-Polyakov-Schwartz-Tyupkin) instanton is an exact solution of Chiral Higher Spin Gravity, i.e. it is also a solution of the holographic dual of Chern-Simons matter theories. This gives an example of a low-spin solution. The instanton sources the opposite helicity spin-one field and a scalar field. We  derive an Effective Field Theory that describes the coupling between an instanton and the other two fields, whose action starts with the Chalmers-Siegel action and has certain higher derivative couplings. 
\end{abstract}

\end{center}
\newpage
\tableofcontents
\newpage
\section{Introduction}
\setcounter{footnote}{0} 
\pagenumbering{arabic}
\setcounter{page}{1}
Higher Spin Gravities (HiSGRA) are theories with massless higher-spin fields, which, as a matter of fact, also include low-spin fields, in particular, the graviton as spin-two (the highest low-spin field), see e.g. \cite{Bekaert:2022poo} for an overview. Masslessness is in a long-term tension with having higher-spin fields in the spectrum, which resolves in having either somewhat peculiar HiSGRAs or somewhat incomplete models.\footnote{This can be argued to be related to the complexity of the quantum gravity problem since any higher spin gravity should be free of UV divergences thanks to the higher symmetry associated with massless higher spin fields. Therefore, such models should not be too easy to construct. It seems that all perturbatively local HiSGRA are consistent with this folklore at present. } Perturbatively local HiSGRAs are: topological $3d$ theories that extend $3d$ (conformal) gravity to higher spins \cite{Blencowe:1988gj, Bergshoeff:1989ns, Campoleoni:2010zq, Henneaux:2010xg, Pope:1989vj, Fradkin:1989xt, Grigoriev:2019xmp, Grigoriev:2020lzu}, conformal HiSGRA \cite{Segal:2002gd, Tseytlin:2002gz, Bekaert:2010ky, Basile:2022nou} and Chiral HiSGRA \cite{Metsaev:1991mt, Metsaev:1991nb, Ponomarev:2016lrm, Skvortsov:2018jea, Skvortsov:2020wtf}. Very close to the latter two is a higher spin extension of IKKT \cite{Sperling:2017dts, Steinacker:2022jjv, Steinacker:2023cuf}.\footnote{Its gauge symmetry is the same as for $4d$ conformal HiSGRA and it features certain truncations that overlap with those of Chiral HiSGRA. It is a non-commutative field theory, though.}

One of the fundamental questions about any theory is the structure of the solution space. The only set-up where constructing exact solutions is not hampered by non-locality and studied in the literature has so far been in three dimensions \cite{Ammon:2011nk,Gutperle:2011kf,Ammon:2012wc,Bunster:2014mua} where the matter-free HiSGRAs can always be formulated as Chern-Simons theories. Therefore, the solutions are characterized by holonomies. Some exact solutions have also been studied for the HS-IKKT model \cite{Sperling:2019xar,Asano:2021phy,Fredenhagen:2021bnw}. 

One of the general features of HiSGRAs is ``spin democracy'', i.e. fields of all spins $s=0,...,\infty$ are equally important members of a single higher spin multiplet. No preference is made even for spin-two, which within general relativity and its low-spin extensions is the one to determine spacetime geometry. A higher spin transformation can activate/deactivate individual spins, e.g. one can nullify any given one. This can lead to confusing effects, e.g. what looks like a black hole metric with a horizon can be mapped to something that is not \cite{Ammon:2011nk}. Therefore, any physical interpretation of solutions should rely on observables that are stable under higher spin transformations (at least the small ones). Such a characterization has been achieved in $3d$ with the help of holonomies. However, extension to higher dimensions is not obvious, but a natural idea is to rely on the invariants of higher spin symmetry \cite{Sezgin:2005pv,Sharapov:2020quq}.  

A related property is that HiSGRAs do not have a dedicated coupling constant to measure the strength at which the higher spins couple to low spins including gravity. As a result, it is not obvious that low-spin solutions can even exist. By default, any spin can serve as a source to any other spin, which is also the case for Chiral HiSGRA. 

There can be several reasons to look for solutions of Chiral theory: (i) this is the only perturbatively local theory with propagating massless fields, i.e. the usual field theory concepts apply; (ii) it should be a consistent truncation of the dual of Chern-Simons matter theories \cite{Sharapov:2022awp} and, hence, all solutions of Chiral theory are also solutions of this bigger yet unknown theory; (iii) the relation to twistors, to self-dual Yang-Mills and self-dual gravity theories as well as the integrability of Chiral theory \cite{Ponomarev:2017nrr,Tran:2021ukl,Tran:2022tft,Herfray:2022prf} should allow for a complete description of the solution space. 

In the present paper we ask whether the famous BPST instanton \cite{Belavin:1975fg} is an exact solution of Chiral theory. The answer is yes and we show how to embed the BPST instanton into $u(2)$-gauged Chiral theory. The spectrum of Chiral theory is given by massless fields of all spins. It can be extended by gauging $u(N)$ Yang-Mills symmetry, after which all fields take values in $u(N)$. Note that there is no $su(N)$-gauging in Chiral theory. Restricting to $N=2$ one can see that the BPST instanton solves one of the Chiral theory's equations provided all higher-spin fields are set to zero. We can assign 'helicity' $+1$ to the instanton. It is clear from the action in the light-cone gauge and from the equations of motion that it should source the $su(2)$-singlet scalar field and an $su(2)$ helicity $-1$ field. Thanks to the $so(4)$ symmetry of the solution all sources to higher-spin fields vanish. 

We also construct a simple Effective Field Theory (low-spin truncation) for $su(2)$ helicity $\pm1$ fields and a singlet scalar that begins with the Chalmers-Siegel action for self-dual Yang-Mills theory \cite{Chalmers:1996rq} and features two higher derivative couplings between them. The exact solution of Chiral HiSGRA we found is also a solution of this simple EFT. The EFT has one coupling constant whose value is fixed by the higher-spin symmetry. 

The outline is as follows. A short introduction into BPST instanton is in Section \ref{sec:BPST}. In Section \ref{sec:EFT} we discuss the EFT and its solutions, which will be shown only later, in Section \ref{sec:BPSTHiSGRA}, to result from Chiral HiSGRA. The self-dual Yang-Mills theory, whose solution the instanton is, is recast into a specific language of Free Differential Algebras in Section \ref{sec:fda}, which is needed to facilitate its embedding into Chiral theory. After a brief overview of the gears of Chiral theory in Section \ref{sec:chiral} we proceed to embedding BPST instanton into the theory in Section \ref{sec:BPSTHiSGRA}. Discussion and conclusions can be found in Section \ref{sec:finale}.

\section{BPST instanton}
\label{sec:BPST}
In this Section we review the BPST instanton \cite{Belavin:1975fg} to recast it into the form most suitable for embedding into chiral higher spin gravity. We will use capital letters from the middle-end of the alphabet $P,Q,R$ to denote the $su(2)$-indices and the usual $A,B,C,\ldots$ and $A',B',C',\ldots$ for the two representations of the Lorentz algebra, which is $sl(2,\mathbb{C})$, $su(2)\oplus su(2)$ and $sl(2,\mathbb{R})\oplus sl(2,\mathbb{R})$ for the Minkowski, Euclidian and split signatures.\footnote{Most of the formulas look the same in all signatures, but the BPST instanton is a solution of the Euclidian Yang-Mills theory.} 

Let $A\equiv A\fdu{P}{Q} \equiv dx^{BB'} A\fdud{P}{Q}{|BB'}$ be an $su(2)$-connection.\footnote{We use the same rules to raise/lower $su(2)$ indices as for the space-time indices, e.g. $v^A=\epsilon^{AB}v_B$, $v_B=v^A\epsilon_{AB}$. If we raise the index on $A\fdu{P}{Q}$ we find $A^{PQ}=A^{QP}$. We sometimes suppress the $su(2)$-indices whenever no confusion can arise. We also use/define $\pl_{AC'}\pl\fdu{B}{C'}=\epsilon_{AB} \square$, i.e. $ \pl_{CC'}\pl^{CC'}=2\square$. The indices in which a tensor is symmetric or that are to be symmetrized are often denoted by the same letter. The symmetrization is defined to be a projector, i.e. one needs to divide by the number of permutations. Lastly, thanks to $A=1,2$ we have $dx^{AA'}\wedge dx^{BB'}=\tfrac12\epsilon^{A'B'} d^2x^{AB}+\tfrac12\epsilon^{AB} d^2x^{A'B'}$, where $d^2x^{AB}\equiv dx\fud{A}{C'}\wedge dx^{BC'}$.} The field strength $F=dA-AA\equiv dA\fdu{P}{R}-A\fdu{P}{Q}\wedge A\fdu{Q}{R}$ can be decomposed into self-dual and anti-self-dual components 
\begin{align}
  F&= F_{AA'|BB'} dx^{AA'}\wedge dx^{BB'}\,, &   F_{AA'|BB'}&= \tfrac12 \epsilon_{AB} F_{A'B'}+\tfrac12 \epsilon_{A'B'} F_{AB}\,,
\end{align}
where $F_{AB}\equiv F\fdu{AC'|B}{C'}$, \textit{idem}. for $F_{A'B'}$. The simplest possibility to get $su(2)$-indices entangled with the spacetime ones reads 
\begin{align}
    A^{PP}= f(x^2) \, x\fud{P}{C'} dx^{PC'}\,.
\end{align}
One can also check that the ansatz satisfies the Lorentz gauge $\partial_{AA'} A^{BB|AA'}=0$. The same can be rewritten as 
\begin{align}
    A^{PP}=\partial\fud{P}{C'} g(x^2)\,\, dx^{PC'}\,,
\end{align}
and it is also convenient to think that $g=\log p$. Thanks to the translation invariance one can replace $x$ with $r=x-a$ in these formulas. We will often use $\xx\equiv r^2\equiv \tfrac12 r_{AA'}r^{AA'}$. This way, in components, we find $A_{PP|AA'}=-\epsilon_{PA}r_{PA'}g'$. For the two terms in the field strength we find
\begin{align}
    dA^{PP}&= \tfrac12 \square g\,d^2x^{PP} -\tfrac12 \pl\fud{P}{M'}\pl\fud{P}{M'}g \,d^2x^{M'M'}\,,\\
    A^{PQ}A\fdu{Q}{P}&= -\tfrac12 (\pl g)^2 \,d^2x^{PP}-\tfrac12 \pl\fud{P}{M'}g \pl\fud{P}{M'}g\, d^2x^{M'M'}\,.
\end{align}
Therefore, the self-duality condition is equivalent to
\begin{align}
    \square g+ (\pl g)^2=0\,,
\end{align}
and with $g = \log p$ becomes simply $\square p=0$. One of the standard choices is 
\begin{align}
    p= 1+\frac{L^2}{r^2}\,,
\end{align}
where $L$ is the size of the instanton. This is usually called a solution in the singular gauge.  
Now, we can check what is left in the field strength
\begin{align}
    F&=\tfrac12 d^2x^{BB}F_{BB}+\tfrac12 d^2x^{B'B'}F_{B'B'}\,.
\end{align}
The anti-self-dual component is the only survivor and it reads
\begin{align}
    F\fud{PP}{|M'M'}&= - \pl\fud{P}{M'}\pl\fud{P}{M'}g+ \pl\fud{P}{M'}g \pl\fud{P}{M'}g=  r\fud{P}{M'}r\fud{P}{M'}\chi_0\,,
\end{align}
where we defined $\chi_0=(-g'' +(g')^2)$. Now, one can compute the Yang-Mills action
\begin{align}
    -\int \Tr[ F_{M'M'} F^{M'M'}]&= \int  F_{PP|M'M'} F^{PP|M'M'}= \int 2\pi r^3\,dr\, 3(r^2)^2 (\chi_0)^2=2\pi\,.
\end{align}
This summarizes what one needs to embed the instanton into Chiral theory.

\section{EFT of BPST instanton from HiSGRA}
\label{sec:EFT}
Chiral theory's spectrum contains massless fields of all spins, i.e. the degrees of freedom that correspond to massless fields with helicities from $-\infty$ to $+\infty$. We consider its $u(2)$-gauged version where all fields take values in $u(2)$. The BPST instanton activates, say, the helicity $+1$ component of the higher spin multiplet, which is associated with $A^{PP}$. We will also find that it induces some source for the helicity $-1$ field and for the scalar field that is an $su(2)$-singlet. Therefore, at least these two fields must not vanish. We can ask what kind of an effective theory (EFT) describes this subsector of Chiral theory. The answer to this question, given in this Section, will be justified later. Let us add also that the EFT here is a truncation of Chiral theory in the sense of dropping all higher-spin fields (including the low-spin sources). It will also have more coupling constants since various interactions can now be considered independent.

Let us start with the Chalmers-Siegel action for self-dual Yang-Mills theory (SDYM). The dynamical fields are: a zero-form $\Psi^{AB}=\Psi^{BA}$ that takes values in the adjoint of $su(2)$, i.e. $\Psi^{AA}\equiv \Psi^{PP|AA}$ if we reveal the $su(2)$-indices; the familiar one-form $su(2)$-connection $A\equiv A^{PP}$. The Lagrangian reads
\begin{align}
    \mathcal{L}[A,\Psi]&= \Tr[\Psi^{AB} F_{AB}(A)]\,,
\end{align}
where $\Tr[ X Y ]\equiv X\fdu{P}{Q} Y\fdu{Q}{P}$ for some $su(2)$-valued $X$ and $Y$. Upon varying with respect to $\Psi^{AB}$ we get the self-duality condition $F_{AB}=0$, whose solutions are instantons. Upon varying with respect to $A$ we get 
\begin{align}
    \nabla\fdu{M}{A'}\Psi^{AM}&=0\,,
\end{align}
where $\nabla=d-A$ is the $su(2)$-covariant derivative, e.g. $[A,\Psi]^{PP}= 2A^{PQ} \Psi\fdu{Q}{P}$. The the Chalmers-Siegel action describes propagation of the helicity $-1$ field over the background created by the positive helicity field $A$. 

The EFT that describes a subsector of Chiral theory relevant for the BPST instanton contains in addition to $\Psi^{AB}$ and $A$ an $su(2)$-singlet scalar field $\phi$. The complete Lagrangian reads
\begin{align}\label{Eftl}
\begin{aligned}
       \mathcal{L}[A,\Psi,\phi]&= \Tr[\Psi^{AB} F_{AB}(A)]+ \tfrac12 \phi \square \phi +\\
       &+\kappa_1 \phi\Tr[F_{A'B'}(A) F^{A'B'}(A)]   + \kappa_2 \Tr[F\fdu{A'}{B'}F^{A'B'}F_{B'B'}] \,.
\end{aligned}
\end{align}
Here $\kappa_{1,2}$ are two coupling constants. Having color indices is very important for the last term to exist. The equations of motion read
\besubeqs\label{eqEFT}
\begin{align}
    F_{AB}(A)&=0\,,\\
    \square \phi&= -\kappa_1\Tr[F_{A'B'} F^{A'B'}]\label{eqScalar}\,,\\
    \nabla\fdu{B}{A'}\Psi^{AB}&= 2 \kappa_1  \nabla^{AB'} \phi F\fud{A'}{B'} + 3\kappa_2  [F_{B'B'}(A), \nabla^{AA'} F^{B'B'}(A)]\,.\label{eqNegaive}
\end{align}
\esubeqs
Note that $\nabla^{AA'} F^{B'B'}(A)$ is symmetric in all primed indices thanks to the Bianchi identities. The commutator, $[\bullet,\bullet]$, means the matrix commutator, $[ X, Y ]\equiv X\fdu{P}{Q} Y\fdu{Q}{R}-Y\fdu{P}{Q} X\fdu{Q}{R}$. If we keep the output indices on the same level and both $X$, $Y$ are in the adjoint of $su(2)$ (no singlet), we get $[ X, Y ]^{PR}\equiv 2X^{(P|Q} Y\fdu{Q}{R)}$. 

If the EFT equations \eqref{eqEFT} are considered non-Lagrangian, then we have three coupling constants, of which two can be fixed at will by rescaling $\Psi$ and $\phi$. Restricting to the Lagrangian case, one cannot change the ratio $\kappa=\kappa_1^2/\kappa_2$, so this is the genuine parameter of the model. In Chiral theory the value of $\kappa$ is fixed by the higher-spin symmetry, but it makes sense to unlock it for the time being.

\paragraph{Solution.} Let us solve the EFT starting with the BPST instanton. The scalar equation acquires the following form
\begin{align}  
    \square \phi&= -\kappa_1\Tr[F_{A'B'} F^{A'B'}]=3\kappa_1 (r^2)^2 (\chi_0)^2\,.
\end{align}
Assuming $\phi=\phi(\xx \equiv r^2)$, the most general solution is
\begin{align}
    \phi&=\frac{2 \kappa _1 L^4}{\xx \left(L^2+\xx\right)^2}-\frac{c_s}{\xx}+c_2\,.
\end{align}
It has especially nice form for $c_{s,2}=0$ and $c_2=0$, $c_s=2\kappa_1$
\begin{align}
    \phi&=\frac{2 \kappa _1 L^4}{\xx \left(L^2+\xx\right)^2}\,, & \phi&= -\frac{2 \kappa _1 \left(2 L^2+\xx\right)}{\left(L^2+\xx\right)^2}\,.
\end{align}
To proceed further, we need
\begin{align}
    \nabla_{MM'}F_{PP|A'A'}&= \pl_{MM'}F_{PP|A'A'}-2 A\fdud{P}{Q}{|MM'} F_{QP|A'A'}= r^{PA'}r^{PA'}r^{MM'} \chi_1\,,
\end{align}
where $\chi_1=\chi_0'-g'\chi_0$ and we used
\begin{align}
    A_{PP|AA^{\prime }}&=-\varepsilon _{PA}r_{PA^{\prime }}g^{\prime }\,.
\end{align}
With $\nabla^{MM'} \phi(\rho\equiv r^2)=r^{MM'} \phi'$ and assuming $\Psi_{PP|AA}=\epsilon_{PA}\epsilon_{PA}f_2(\rho)$ we get for \eqref{eqNegaive}
\begin{align}
    \epsilon^{PA}r^{PA'}&: && f_2'-2g'f_2=2 \kappa_1 r^2 \phi' \chi_0 -4\kappa_2 (r^2)^2 \chi_0\chi_1 \,,
\end{align}
which is a scalar equation since all terms feature the same spin-tensor structure displayed on the left. The general solution reads
\begin{align}\label{genneg}
    f_2&=\frac{4 \kappa _1 c_s L^2}{3 \xx^2 \left(L^2+\xx\right)}-\frac{8 \kappa _1^2 L^6 \left(4 L^2+9 \xx\right)}{15 \xx^2 \left(L^2+\xx\right)^4}-\frac{4 \kappa _2 L^4 \left(6 L^2 \xx+L^4+15 \xx^2\right)}{5 \xx^2 \left(L^2+\xx\right)^4}+\frac{c_1 \left(L^2+\xx\right)^2}{\xx^2}\,.
\end{align}
There is a special point in the parameter space, $c_1=0$, $c_s=2\kappa_1$, $\kappa_2=\kappa_1^2/3$, where the solution drastically simplifies:
\begin{align}
    f_2&=\frac{8 \kappa _1^2 L^2 \xx}{3 \left(L^2+\xx\right)^4}\,.
\end{align}

\paragraph{Light-cone gauge/spinor-helicity.} It is not hard to derive this EFT, \eqref{Eftl}, by looking at the action of Chiral theory in the light-cone gauge and assuming that only helicities $0$, $\pm1$ can participate. In flat space and in the light-cone gauge the action has cubic interactions only and reads
\begin{align}\label{LCaction}
    S&= \sum_{s\geq0}\int \Phi_{-s} \square \Phi_s + \sum_{\lambda_{1,2,3}}\frac{l_p^{\lambda_1+\lambda_2+\lambda_3-1}}{\Gamma[\lambda_1+\lambda_2+\lambda_3]}\int  V_{\lambda_1,\lambda_2,\lambda_3} \Phi_{\lambda_1}\Phi_{\lambda_2}\Phi_{\lambda_3}\,,
\end{align}
where $\Phi_\lambda$ is a 'scalar' representing the helicity $\lambda$ degree of freedom. Here, $l_p$ is a coupling constant with the dimension of length. On-shell the vertices reduce to the well-known spinor-helicity expression
\begin{align}\label{genericV}
   V_{\lambda_1,\lambda_2,\lambda_3}\Big|_{\text{on-shell}} \sim 
    [12]^{\lambda_1+\lambda_2-\lambda_3}[23]^{\lambda_2+\lambda_3-\lambda_1}[13]^{\lambda_1+\lambda_3-\lambda_2}\,.
\end{align}
Restricting to $\pm 1$ and $0$ subsector we find the following in \eqref{genericV} that matches \eqref{Eftl}. The vertices must have the total helicity positive and, hence, we can have $V_{0,0,1}$, $V_{0,1,1}$, $V_{1,1,-1}$ and $V_{1,1,1}$. The first option, $V_{0,0,1}$, which is the current interaction, cannot be realized for a singlet scalar field. $V_{1,1,-1}$ is the half of the Yang-Mills vertex that is present in SDYM, it is captured by the cubic part of the first term in $\mathcal{L}$ \eqref{Eftl}. $V_{0,1,1}$ is the $\kappa_1$-term and $V_{1,1,1}$ is the $\kappa_2$-term. Note that the kinetic term of the Chalmers-Siegel action contracts $+1$ to $-1$, i.e. the equation for $\Psi$ are obtained by varying with respect to $A$ and other way around. Also, sum of the helicities in a vertex is equal to the number of derivatives in the corresponding covariant vertex. 

It is important to realize that low-spin fields do source higher-spin fields, in general. The sources, as we will show later, vanish on the BPST instanton, but do not have to vanish on other solutions of the EFT. It would be interesting to probe the EFT with a generic ADHM-instanton \cite{Atiyah:1978ri}.

\section{BPST as a Free Differential Algebra}
\label{sec:fda}
Chiral theory's covariant equations of motion are formulated as a Free Differential Algebra (FDA) \cite{Sullivan77,vanNieuwenhuizen:1982zf,DAuria:1980cmy,Vasiliev:1980as} or in the AKSZ-form \cite{Alexandrov:1995kv}. Any (gauge) theory can be written in such a form \cite{Barnich:2010sw,Grigoriev:2012xg}, but usually it is not needed and is also hard to do explicitly. The price to pay is to introduce infinitely-many auxiliary fields. The FDA for SDYM was found in \cite{Skvortsov:2022unu} and we repeat some of the main steps in the derivation to make the paper self-contained.\footnote{Another example is the self-interacting scalar field  \cite{Misuna:2022cma,Misuna:2024ccj}. }

As the first step the self-duality condition can be rewritten as\footnote{The numerical factor of $-1/2$ does not have any physical significance and is only there to agree with what emerges from Chiral theory.}
\begin{align}
    dA^{PP}-A^{PM}\wedge A\fdu{M}{P}= -\tfrac12 dx_{BA'} \wedge dx\fud{B}{A'}C^{PP|A'A'}\,.
\end{align}
It implies that $F_{AB}=0$, but has an auxiliary fields $C_{A'A'}$ as a plug to account for the fact that $F_{A'B'}$ does not have to vanish. Basically, $C_{A'A'}\sim F_{A'A'}$, to be precise $C_{A'A'}=- F_{A'A'}$. As a result, $C_{A'A'}$ satisfies the Bianchi identities $\nabla\fud{A}{M'}C^{A'M'}=0$, which can be reformulated in a more positive way as
\begin{align}
    \nabla C^{A'A'}&= dx_{MM'} C^{M,A'A'M'}\,,
\end{align}
where $C^{A,A'A'A'}$ is an irreducible spin-tensor, as the notation suggests. This is a consequence of the Bianchi identity. One can hit the last formula with $\nabla$ to get a constraint on $\nabla C^{A,A'A'A'}$, which can be solved with the help of another auxiliary field and so on. Starting from this point the equations become nonlinear, but the nonlinearities do not get worse than bilinear in the fields $C$. The final result of this procedure \cite{Skvortsov:2022unu} reads (see Appendix \ref{app:sdym} for more details)
\begin{align}
dC_{A(k),A^{\prime }(k+2)} &=\left[ A,C_{A(k),A^{\prime }(k+2)}\right]
+dx^{BB^{\prime }}C_{A(k)B,A^{\prime }(k+2)B^{\prime }}  \notag \\
&+\sum_{n=0}^{k-1}\tfrac{\left( k+2\right) !}{\left( n+1\right) !\left(
k-n-1\right) !\left( k+1\right) \left( k-n+1\right) }dx_{A}{}^{B^{\prime }}%
\left[ C_{A(n),A^{\prime }(n+1)B^{\prime }},C_{A(k-n-1),A^{\prime }(k-n+1)}%
\right] \,.  \label{dCeqACetc}
\end{align}
The complete set of fields is thereby given by $A$ and $C_{A(k),A'(k+2)}$. Sticking to the free limit, i.e. dropping the bilinear terms, we simply get
\begin{align}
dC_{A(k),A^{\prime }(k+2)} &=dx^{BB^{\prime }}C_{A(k)B,A^{\prime }(k+2)B^{\prime }} 
\end{align}
or, in components,
\begin{align}
\pl_{MM'}C_{A(k),A^{\prime }(k+2)} &=C_{A(k)M,A^{\prime }(k+2)M^{\prime }} \,.
\end{align}
Therefore, the $k>0$ auxiliary fields parameterize higher derivatives of the anti-self-dual field strength $C_{A'A'}$, i.e. $C_{A(k),A'(k+2)}=\pl_{AA'}...\pl_{AA'}C_{A'A'}$, which, hopefully demystifies the specific choice of auxiliary fields. All other derivatives of $C_{A'A'}$, e.g. $\square C_{A'A'}$, vanish on-shell.

In order to solve (\ref{dCeqACetc}) we rewrite it as components for $%
dx^{BB^{\prime }}$:%
\begin{align}
\partial _{BB^{\prime }}C_{CC|A(k),A^{\prime }(k+2)}  \notag 
&=2A\fdud{C}{D}{|BB^{\prime }}C_{CD|A(k),A^{\prime
}(k+2)}+C_{CC|A(k)B,A^{\prime }(k+2)B^{\prime }}  \label{dCeq1} \\
&+2\sum_{n=0}^{k-1}\tfrac{\left( k+2\right) !}{\left( n+1\right) !\left(
k-n-1\right) !\left( k+1\right) \left( k-n+1\right) }\varepsilon
_{BA}C\fdud{C}{D}{|A(n),A^{\prime }(n+1)B^{\prime }}C_{CD|A(k-n-1),A^{\prime
}(k-n+1)}\,.\notag
\end{align}
At this point it is useful to introduce a generating function for all $C_{PP|A(k),A'(k+2)}$ as
\begin{equation}
C_{PP}=\sum\limits_{k=0}^{\infty }\frac{l^{k+2}}{%
k!\left( k+2\right) !}C_{PP|A(k),A^{\prime }(k+2)}\, y^{A}...y^A\,\bar{y}^{A^{\prime }}...\bry^{A'}\,,  \label{solcomp}
\end{equation}%
where $l$ has the unit of length to account for the fact that the $k$-th field is the $k$-th order derivative of $C_{PP|A'A'}$. With the help of the symmetries of the BPST instanton it is easy to conclude that
\begin{equation}
C_{AA|A(k),A^{\prime }(k+2)}=\left( k+2\right) !\left( r_{AA^{\prime
}}\right) ^{k+2}h_{k}\ .
\end{equation}%
for some $h_k(r^2)$. The over-determined system for $h_k$ that follows from (\ref{dCeqACetc}) can be solved, see Appendix \ref{app:sdym}, to give
\begin{equation}
C_{AA|A(k),A^{\prime }(k+2)}=\left( r_{AA^{\prime }}\right) ^{k+2}\left(
-1\right) ^{k+1}\left( k+2\right) !\frac{L^{2}}{r^{2}\left(
L^{2}+r^{2}\right) ^{k+2}}\ ,  \label{Ckarb}
\end{equation}%
and, hence, the generating function reads
\begin{eqnarray}
C_{PP} &=\frac{-l^{2}L^{2}}{r^{2}\left( L^{2}+r^{2}\right) ^{2}}\left(
r_{PB^{\prime }}\bar{y}^{B^{\prime }}\right)  \left(
r_{PC^{\prime }}\bar{y}^{C^{\prime }}\right)e^{\frac{-l}{L^{2}+r^{2}}%
r_{BB^{\prime }}y^{B}\bar{y}^{B^{\prime }}}\ .
\end{eqnarray}
A simpler way of solving this system will be presented in Section \ref{sec:BPSTHiSGRA} and relies on some higher-spin techniques, which can, in principle, be avoided here. 

\section{Chiral Higher Spin Gravity}
\label{sec:chiral}
To begin with, the main difference is that the standard approach to higher spin fields where the main field variables are Fronsdal fields, $\Phi_{\mu_1...\mu_s}$, should be replaced with the chiral description that originates naturally from twistor theory \cite{Penrose:1965am,Hitchin:1980hp}. In the latter approach positive and negative helicities are treated in a different way. Free massless fields with $s>0$ require a one-form $\omega^{A(2s-2)}\equiv \omega^{A(2s-2)}_\mu \,dx^\mu$ that is a totally symmetric rank-$(2s-2)$ spin-tensor of the Lorentz algebra and a zero-form $\Psi^{A(2s)}$ that is a symmetric rank-$2s$ spin-tensor. The free action can be thought of as a straightforward generalization of the Chalmers-Siegel action \cite{Chalmers:1996rq} to higher spins and it reads \cite{Krasnov:2021nsq}
\begin{align}\label{niceaction}
    S= \int \Psi^{A(2s)}\wedge e_{AB'}\wedge e\fdu{A}{B'}\wedge \nabla \omega_{A(2s-2)}\,,
\end{align}
where $e^{AA'}\equiv e^{AA'}_\mu \, dx^\mu$ is the background vierbein. The action enjoys a gauge symmetry
\begin{align}\label{lin-gauge}
    \delta \omega^{A(2s-2)}&= \nabla \xi^{A(2s-2)} +e\fud{A}{C'} \eta^{A(2s-3),C'}\,,& \delta\Psi^{A(2s)}&=0\,,
\end{align}
where $\xi^{A(2s-2)}$ and $\eta^{A(2s-3),C'}$ are zero-forms. $\nabla$ is the Lorentz covariant derivative on any self-dual background. The equations of motion obtained from \eqref{niceaction} read
\begin{align}\label{first}
    \nabla \Psi^{A(2s)}\wedge H_{AA}&=0\,, && H^{AA}\wedge \nabla \omega^{A(2s-2)}=0\,,
\end{align}
where we defined $H_{AA}=e_{AB'}\wedge e\fdu{A}{B'}$.

\paragraph{Free FDA.} The equations of motion resulting from the action \eqref{niceaction} can be rewritten in the FDA form, which requires an appropriate set of auxiliary fields. The latter can be packaged into two generating functions\footnote{This set of fields was proposed in \cite{Vasiliev:1986td} to describe FDA of Fronsdal fields. Even though the dynamical fields and the equations are different now, the set of auxiliary fields covers the same space, which is not surprising since it is determined by the physical degrees of freedom.}
\begin{align*}
    \omega(y,\bry)&= \sum_{n+m=\text{even}}\tfrac{1}{n!m!} \omega_{A(n),A'(m)}\, y^A...y^A\, \bry^{A'}...\bry^{A'} \,,
 \end{align*}  
for one-forms and 
\begin{align*}
 C(y,\bry)&= \sum_{n+m=\text{even}}\tfrac{1}{n!m!} C_{A(n),A'(m)}\, y^A...y^A\, \bry^{A'}...\bry^{A'}\,,
\end{align*}
for zero-forms. The dynamical fields $\Psi^{A(2s)}$ (together with the scalar field $\phi=\Psi^{A(0)}$), $\omega^{A(2s-2)}$ are identified with $C(y,\bry=0)$ and $\omega(y,\bry=0)$, respectively. The free equations read \cite{Skvortsov:2022syz}
\besubeqs\label{linearizeddata}
\begin{align}
    \nabla\omega &= -2e^{BB'}y_{B} \plb_{B'}\, \omega -2e\fdu{A}{B'}\wedge e^{AB'} \plb_{B'}\plb_{B'}C(y=0,\bry)\,,\\
    \nabla C&= 2 e^{BB'}\pl_B \plb_{B'}\, C\,,
\end{align}
\esubeqs
where $\nabla e^{AA'}=0$ is the background vierbein and $\nabla^2=0$, i.e. it describes the flat space. Note that the fluctuation $\omega$ does also contain the spin-two sector, which includes a perturbation of the vierbein, $\omega_{A,A'}$. To some extent the Fronsdal fields are still present in the system and are associated with the totally-symmetric component of the higher-spin vierbein
\begin{align}
    \Phi_{\mu_1...\mu_s}&= \omega^{\ga(s-1),\gad(s-1)}_{(\mu_1} \, e_{\mu_2|\ga\gad}...e_{\mu_s)|\ga\gad}\,.
\end{align}
The following diagram shows how fields 'talk' to each other at the free level
\begin{figure}[h!]
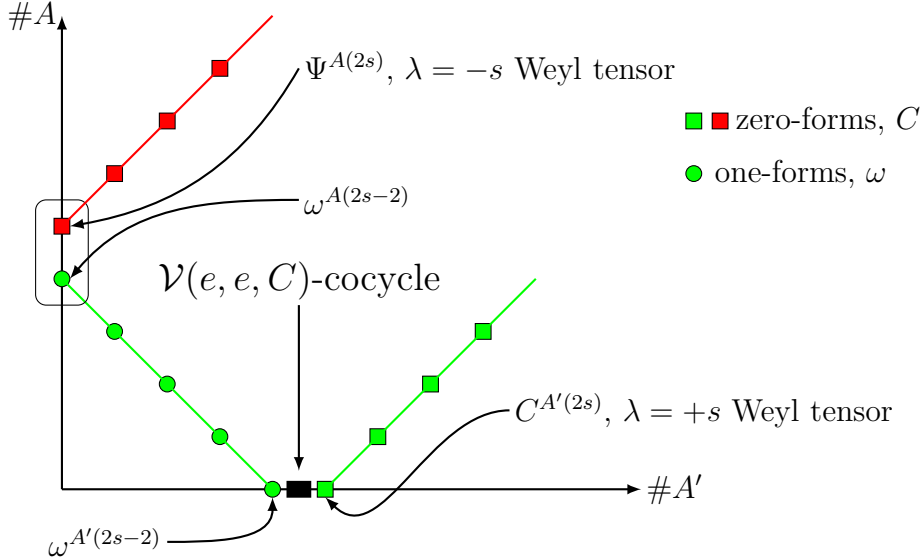

\FIELDS
  \caption{The picture illustrates where positive/negative helicity fields reside together with their auxiliary fields. The dynamical fields are zero form $\Psi^{A(2s)}$ and one-form $\omega^{A(2s-2)}$. The former has an infinite set of zero-form auxiliary fields denoted by the red boxes. The latter has a finite set of auxiliary one-forms denoted by the green bullets that is later joined by an infinite set of zero-forms denoted by the green boxes. Conventionally, positive helicity fields, $\lambda=+s$, are shown in green and the negative helicity fields are shown in red. The links between fields are due to the free equations of motion that relate derivative of one to another one. }\label{fig:figure1}
\end{figure}

Interactions correspond to certain deformations of these free equations that are formally consistent (hence, gauge invariant) and obey locality (hence, make sense). The most general ansatz reads
\besubeqs\label{eq:chiraltheory}
\begin{align} 
    d\omega&= \mathcal{V}(\omega, \omega) +\mathcal{V}(\omega,\omega,C)+\mathcal{V}(\omega,\omega,C,C)+...\,,\\
    dC&= \mathcal{U}(\omega,C)+ \mathcal{U}(\omega,C,C)+... \,.
\end{align}
\esubeqs
Here, we assumed that the free equations \eqref{linearizeddata} result from linearization of the equations above, hence, for example, the background vierbein $e^{AA'}$ can be absorbed into the full master-field $\omega$. The free equations \eqref{linearizeddata} impose certain boundary conditions on the first three vertices:
\besubeqs\label{eq:boundaryconditions}
\begin{align}
    \mathcal{V}(e,\omega)+\mathcal{V}(\omega,e)&\sim e^{BB'} y_{B} \plb_{B'}\omega\,,\\
    \mathcal{U}(e,C)+\mathcal{U}(C,e)&\sim e^{BB'} \pl_B \plb_{B'}  C\,,\\
    \mathcal{V}(e,e,C)& \sim e\fdu{B}{C'}e^{BC'} \pl_{C'} \pl_{C'} C(y=0,\bry)\,.
\end{align}
\esubeqs
Here, $e=e^{AA'}\, y_A \bry_{A'}$. In order to systematically introduce interactions we need the efficient language of symbols of operators. 

\paragraph{Poly-differential operators.} Vertices $\mathcal{V}$ and $\mathcal{U}$ encode certain contractions of indices of their arguments, e.g.
\begin{align}
    \mathcal{V}(\omega,\omega,C,...,C)= \sum y_A...y_A \,\omega\fud{A(...)}{B(...)M(...) ...}\wedge \omega\fud{A(...)B(...)}{N(...)} C^{A(...)N(...) ...} ... \,,
\end{align}
where we omitted $\bry$. It is convenient to represent such structures via poly-differential operators
\begin{align}
    \mathcal{V}(f_1,...,f_n)&= \mathcal{V}(y, \pl_1,...,\pl_n)\, f_1(y_1)...f_n(y_n) \Big|_{y_i=0}\,.
\end{align}
We prefer to work with the corresponding symbols, obtained by replacing  the arguments according to $y^A\equiv p_0^A$, $\pl^{y_i}_{A}\equiv p_{A}^i$. The Lorentz symmetry requires the symbols to depend only on $p_{ij}\equiv p_i \cdot p_j\equiv -\epsilon_{AB}p^A_{i}p_{j}^B=p^A_{i}p_{jA}$. These scalars are defined so that $\exp[p_0\cdot p_i]f(y_i)=f(y_i+y)$ represents the translation operator. We will also use $q$'s for poly-differential operators in $\bry$, e.g. $\bry^{A'}\equiv q_0^{A'}$, $\pl^{\bry_i}_{A'}\equiv q_{A'}^i$. We usually omit $|_{y_i=0}$ and the arguments of the vertices, writing down only the corresponding symbols. 

For example, the boundary conditions \eqref{eq:boundaryconditions} can be rewritten in the full form as 
\besubeqs\label{eq:boundaryconditionsB}
\begin{align}
    \mathcal{V}(e,\omega)+\mathcal{V}(\omega,e)&\sim (p_{01}q_{12})\, e^{p_{02}+q_{02}}\, ( e^{CC'}y^1_C \bry^1_{C'})\, \omega(y_2,\bry_2)\Big|_{y_{1,2}=\bry_{1,2}=0}\label{eq:boundaryconditionsBA}\,,\\
    \mathcal{U}(e,C)+\mathcal{U}(C,e)&\sim (q_{12} p_{12})\, e^{p_{02}+q_{02}}\, ( e^{CC'}y^1_C \bry^1_{C'})\, C(y_2,\bry_2)\Big|_{y_{1,2}=\bry_{1,2}=0}\label{eq:boundaryconditionsBB}\,,\\
    \mathcal{V}(e,e,C)&\sim  q_{13} q_{23} p_{12}\, e^{q_{03}}\,( e^{BB'}y^1_B \bry^1_{B'})( e^{CC'}y^2_C \bry^2_{C'})\, C(y_3,\bry_3)\Big|_{y_{1,2,3}=\bry_{1,2,3}=0}\label{eq:boundaryconditionsBC}\,,
\end{align}
\esubeqs
where $\sim$ implies an unessential numerical coefficient. Here we used the fact that there are no matrix factors (see below) to bring several components of each vertex into the same ordering. In what follows we usually display the symbols of operators, but not the arguments.

\paragraph{Higher Spin Algebra.} The first bilinear map defines the higher spin algebra $\hs$
\begin{align}\label{hsalgebra}
    \mathcal{V}(f,g)&= \exp{[p_{01}+p_{02}]}\exp{[q_{01}+q_{02}+q_{12}]}f({y}_1)\, g({y}_2)\Big|_{{y}_i=0} \equiv (f\star g)(y)\,.
\end{align}
This is just the star-product in $\bry$ and the commutative product in $y$. 

One feature of Chiral FDA is that we always have just star-product over $\bry$ variables. More generally, one can think of fields $\omega$, $C$ as taking values in $\mathbb{C}[y] \otimes B$, where $B$ is some associative algebra. In other words, all vertices have a factorized form
\begin{align}
    \mathcal{V}(f_1,...,f_n)&= v(f_1'(y),..., f_n'(y)) \otimes f_1''\star... \star f_n''  \,,
\end{align}
where $f_i=f_i'(y) \otimes f_i''$, $f''_i\in B$. An option to enrich the vertices by any associative algebra is thanks to the underlying structure being an $A_\infty$-algebra, but we will not need any further details. For ungauged Chiral theory we choose $B=A_1[\bry]$, where $A_1$ is the Weyl algebra. As is well-known, $A_1$ can be realized as functions in $\bry$ equipped with the Moyal-Weyl star product, as above. Yang-Mills gaugings can be added via an additional matrix factor, $B=A_1[\bry]\otimes \mathrm{Mat}_M$. One can also add supersymmetry via a factor of Clifford algebra, see \cite{Metsaev:2019dqt,Metsaev:2019aig,Tsulaia:2022csz}. 

\paragraph{The dual module.} The next bilinear vertex splits into two vertices
\begin{align}
    \mathcal{U}(\omega,C)&=\mathcal{U}_1(\omega,C)+\mathcal{U}_2(C,\omega)\,.
\end{align}
The $A_\infty$-relations imply that zero-forms take values in some bimodule of $\hs$. It turns out that the zero-forms take values in the dual module. To define the dual action we need to pick some non-degenerate bilinear form
\begin{align}
    \langle a|u \rangle&= \exp[p_{12}]\,a(y_1)\,u(y_2) \big|_{y_i=0}\,.
\end{align}
between an $\hs$ bi-module $M$ where $C$ takes values and higher spin algebra $\hs$. Then, the dual module action reads
\begin{equation}
    \begin{split}
        &\mathcal{U}_1(\omega,C)=+\exp{[ p_{02}+p_{12}]}\exp{[q_{01}+q_{02}+q_{12}]}\, \omega({y}_1)\, C({y}_2)\Big|_{y_i, \bar{y}_i=0}\,,\\
        &\mathcal{U}_2(C,\omega)=-\exp{[p_{01}-p_{12}]}\exp{[q_{01}+q_{02}+q_{12}]}\, C({y}_1)\, \omega({y}_2)\Big|_{y_i, \bar{y}_i=0}\,.
    \end{split}
\end{equation}
We note that we consider the bosonic theory, i.e. $\omega$ and $C$ are even functions. $\mathcal{U}(\omega,C)$ is just the action of the commutative algebra on the dual space $\mathcal{U}_1(\omega,C)(y)=\omega(\pl_y) C(y)$, which is by differential operators. 

\paragraph{$\boldsymbol{\mathcal{V}(\omega,\omega,C)}$.} We will not need vertices beyond the cubic ones, see Section \ref{subsec:decoupling}, and those can be just written down directly. There are $3$ structure maps hidden in $\mathcal{V}(\omega,\omega,C)$\footnote{We slightly abuse notation here: since the structure maps are $A_\infty$-maps, the order of the arguments is important (in case we have Yang-Mills gaugings we cannot permute $\omega$ and $C$ at all). Here, $\mathcal{V}(\omega,\omega,C)$ is just a shorthand notation for all $\omega^2C$-type vertices, while the individual structure maps with specific ordering of arguments are displayed on the right. }
\begin{align}
    \mathcal{V}(\omega,\omega,C)=\mathcal{V}_1(\omega,\omega,C)+\mathcal{V}_2(\omega,C,\omega)+\mathcal{V}_3(C,\omega,\omega)    \,.
\end{align}
They have a very simple form
\begin{align*}
     \mathcal{V}_1(\omega,\omega,C)&=+p_{12}\, \int_{\Delta_2}\exp[\left(1-t_1\right) p_{01}+\left(1-t_2\right) p_{02}+t_1 p_{13}+t_2 p_{23} ]\,, \\
     \mathcal{V}_2(\omega,C,\omega)&=-p_{13}\, \int_{\Delta_2}\exp[\left(1-t_2\right) p_{01}+\left(1-t_1\right) p_{03}+t_2 p_{12}-t_1 p_{23}]\\
       &\phantom{=}\,-p_{13}\, \int_{\Delta_2}\exp[\left(1-t_1\right) p_{01}+\left(1-t_2\right) p_{03}+t_1 p_{12}-t_2 p_{23}]\,,
    \\
    \mathcal{V}_3(C,\omega,\omega)&=+p_{23}\, \int_{\Delta_2}\exp[\left(1-t_2\right) p_{02}+\left(1-t_1\right) p_{03}-t_2 p_{12}-t_1 p_{13} ]\,.
\end{align*}
where $\Delta_2$ is the simplex $0<t_1<t_2<1$. We have also dropped here the star-product factor over $\bry$, which is 
\begin{align}
    \exp{[q_{01}+q_{02}+q_{03}+q_{12}+q_{13}+q_{23}]}\,.
\end{align}
If $u(N)$-symmetry is gauged, $\omega$ and $C$ are considered as matrix-valued fields, $\omega\equiv \omega(y,\bry)\fdu{P}{R}$, $C\equiv C(y,\bry)\fdu{P}{R}$, where the matrix factors are multiplied in the way the arguments of the vertices are written. 

\paragraph{$\boldsymbol{\mathcal{U}(\omega,C,C)}$.} Similarly, we have three maps for the second type of cubic vertices
\begin{align}
    \mathcal{U}(\omega,C,C)=\mathcal{U}_1(\omega,C,C)+\mathcal{U}_2(C,\omega,C)+\mathcal{U}_3(C,C,\omega)   \,.
\end{align}
These maps are not independent of $\mathcal{V}(\omega,
\omega,C)$: the $A_\infty$-algebra turns out to be of pre-Calabi--Yau type, which means, in practice, that many structure maps are related via certain duality. For example, we find
\besubeqs
\begin{align}
    \mathcal{U}_1(p_0,p_1,p_2,p_3)&=+ \mathcal{V}_1(-p_3,p_0,p_1,p_2)\,,\\
    \mathcal{U}_2(p_0,p_1,p_2,p_3)&=- \mathcal{V}_2(-p_1,p_2,p_3,p_0)\,,\\
    \mathcal{U}_3(p_0,p_1,p_2,p_3)&= -\mathcal{V}_3(-p_1,p_2,p_3,p_0)\,.
\end{align}
\esubeqs
Explicitly we have 
\begin{align}
    \mathcal{U}_1&=p_{01} \int_{\Delta_2}\exp \left[t_1 p_{02}+\left(1-t_1\right) p_{03}+t_2 p_{12}+\left(1-t_2\right) p_{13}\right]\,,\\
    \mathcal{U}_2&=-p_{02} \int_{\Delta_2}\exp \left[\left(1-t_2\right) p_{01}+t_2 p_{03}-\left(1-t_1\right) p_{12}+t_1 p_{23}\right]\\
    &\phantom{=}-p_{02} \int_{\Delta_2}\exp \left[\left(1-t_1\right) p_{01}+t_1 p_{03}-\left(1-t_2\right) p_{12}+t_2 p_{23}\right]\,,\\
    \mathcal{U}_3&= +p_{03} \int_{\Delta_2}\exp \left[\left(1-t_1\right) p_{01}+t_1 p_{02}-\left(1-t_2\right) p_{13}-t_2 p_{23}\right]\,.
\end{align}
Note that, for instance, there is no $p_{23}$ in $\mathcal{U}_1$, which implies locality.

\subsection{Comments on higher orders}
\label{sec:}
Let us discuss the general structure of higher order vertices, which will help later, in Section \ref{subsec:decoupling}, to explain why they are not needed. Local vertices of Chiral HiSGRA have a very special form. The $\mathcal{V}$-vertices read
\begin{align}\label{higherV}
    \mathcal{V}(\omega,\omega,C,...,C)&= (p_{12})^n \exp[\ast p_{01}+\ast p_{02}+ \sum_{2<i\leq n+2} \ast p_{1i}+\sum_{2<i\leq n+2} \ast p_{2i}]\,,
\end{align}
where $n$ is the number of zero-form arguments $C$. 
Here $\ast$ denote some functions of the 'times' $t_i$ that are integrated over a compact domain. 
The prefactor $p_{12}^n$ means that all higher order vertices vanish for low-spin $\omega$. An interesting effect is that for every fixed spin in $\omega$ there is always some maximal order where its contribution stops. Therefore, it might be possible to have a class of solutions where $\omega$ has a bounded number of spins activated. The $\mathcal{U}$-vertices are obtained via the duality map
\begin{align*}\label{higherU}
   &\mathcal{U}(p_0,p_1,...,p_{n+2}) =\mathcal{V}(-p_{n+2},p_0,p_1,...,p_{n+1}) = \\
  & (p_{01})^n \exp[ \ast p_{0,n+2}+\ast p_{1,n+2}+\sum_{1<i\leq n+1} \ast p_{0,i}+\sum_{1<i\leq n+1} \ast p_{1,i}]\,.
\end{align*}
The locality is encoded in the fact that there are no $p_{ij}$ in the exponent that contract indices on any two zero-forms. It should be remembered that there is always $\star$-star product in $\bry$, which is implicit. At the free level it is clear that auxiliary fields express higher derivatives in the form of $y^A \partial_{AA'} \bry^{A'}$. Therefore, having $p_{ij}$ ($i$, $j$ connect zero-forms) in the exponent would imply nonlocality since we already have $q_{ij}$ due to the star-product in $\bry$. Indeed, taking a derivative $\pl_{AA'}$ produces a pair of indices contracted with $y^A\bry^{A'}$ in generating function $C$. Given that there is $\exp[q_{23}]$ already present in the vertex, it is easy to see that $\exp[\ast p_{23} +q_{23}]$ will produce an infinite series of contracted derivatives, i.e. such a vertex is non-local. 

\subsection{Low-spin (de)coupling}
\label{subsec:decoupling}
Let us consider the $u(2)$-gauged Chiral theory. In this theory $\omega=\omega(y,\bry)\fdu{P}{Q}$ and $C=C(y,\bry)\fdu{P}{Q}$. Let us further consider solutions that activate only the low-spin subsector of Chiral theory, i.e. we have the following potentially nonvanishing components of $\omega$ and $C$:
\besubeqs\label{lowspinan}
\begin{align}
    \omega^{PQ}&= A^{PQ} + \epsilon^{PQ} \big[\tfrac12 \omega^{AA} y_A y_A + \tfrac12 \omega^{A'A'} \bry_{A'} \bry_{A'} + e^{AA'} y_A \bry_{A'}\big]\,,\\
    C^{PQ}&= \sum_{s=0,1,2}\sum_{k} C\fud{PQ|}{A(k),A'(2s+k)}\, y^{A(k)}\, \bry^{A'(2s+k)}+C\fud{PQ|}{A(2s+k),A'(k)}\, y^{A(2s+k)}\, \bry^{A'(k)}\,,
\end{align}
\esubeqs
where in the last line we did not include the factorials as we will work with the generating functions as the whole. 
The gravitational subsector of $\omega$ and $C$ must be an $su(2)$-singlet, i.e. is a multiple of $\epsilon^{PQ}$. In practice, we will see that the spin-two will be in its vacuum, i.e. the corresponding components of $C$ vanish. 

A simple calculation with the lower order vertices evaluated on the low-spin ansatz with non-abelian $A$ and $C$, but abelian gravitational sector, leads to
\besubeqs\label{lowspineqs}
\begin{align}
    dA&= AA -2e_{MB'}\wedge e\fud{M}{B'} C^{B'B'}\,,\label{eqSD}\\
    d\omega^{AA}&= -2 e\fud{A}{C'}\wedge e^{AC'}\,,\label{eqOmega}\\
    de^{AA'}&= -\omega\fud{A'}{C'}\wedge e^{AC'}\,,\label{eqTorsion}\\
    d\omega^{A'A'}&= -2\omega\fud{A'}{C'}\wedge \omega^{A'C'} -2 e_{MB'}\wedge e\fud{M}{B'} C^{A'A'B'B'}\,,\label{eqEinstein}\\
    dC&= [A,C]-\omega^{A'A'}\bry_{A'}\pl_{A'} C +2e^{AA'}\pl_A \pl_{A'}C+ \mathcal{U}(e,C,C)\,,\label{eqC}
\end{align}
\esubeqs
where $[A,C]_{PR}\equiv A\fdu{P}{Q} C_{QR}+A\fdu{R}{Q} C_{PQ}$. The source $\mathcal{U}(e,C,C)$ has the following form
\begin{align*}
    \mathcal{U}(e,C,C)= p_{01} e^{q_{02}+q_{03}+q_{23}} \int\Big( &q_{12} e^{t_1 p_{02}-t_1 p_{03}+p_{03}}+q_{12} e^{-t_2 p_{02}+p_{02}+t_2 p_{03}}\\
     -&q_{13} e^{t_2 p_{02}-t_2 p_{03}+p_{03}}-q_{13} e^{-t_1 p_{02}+p_{02}+t_1 p_{03}}\Big)e(1) C(2) C(3)\,,
\end{align*}
where $e(1)$, $C(2,3)$ means, e.g. $C=C(y_2,\bry_2)$, and $|_{y_i=\bry_i=0}$ is omitted. In deriving this expression we moved $e$ to the first argument since it does not get entangled with the color indices. $\mathbb{Z}_2$-symmetry of the integration domain was also used, i.e. one can replace $t_1\rightarrow1-t_2$, $t_2\rightarrow 1-t_1$. In a less symmetric but more compact form we have
\begin{align}
        \mathcal{U}(e,C,C)&= 2\,p_{01} e^{q_{02}+q_{03}+q_{23}} \int\Big( q_{12} e^{t_1 p_{02}+(1-t_1) p_{03}}-q_{13} e^{(1-t_1) p_{02}+t_1 p_{03}}\Big)e(1) C(2) C(3)\,.
\end{align}
If we drop $\exp{q_{23}}$ then the expression is antisymmetric under $2\leftrightarrow 3$ swap. Let us make few comments to explain the structure of the equations. Eq. \eqref{eqSD} imposes the familiar self-duality relation on $A$. Eq. \eqref{eqOmega} tells us that $\omega^{AA}$ is not a (half of) spin-connection, but it is the right variable to be a dynamical field in self-dual gravity \cite{Krasnov:2016emc}. Nevertheless, $\omega^{A'A'}$ does behave as (half of) the spin-connection and Eq. \eqref{eqTorsion} is a torsion constraint. Eq. \eqref{eqEinstein} sets the self-dual component of the Weyl tensor to zero and imposes Einstein equations. Similarly to \eqref{eqSD}, the nonvanishing component of the Riemann tensor is encoded in $C_{A'A'A'A'}$. The equations of motion for the negative helicity fields $\Psi^{A(2s)}$ and Bianchi identities for $C_{A'A'}$ and $C_{A'A'A'A'}$ are hidden in Eq. \eqref{eqC}. The equation for the scalar field is also in Eq. \eqref{eqC}.

\paragraph{Light-cone glasses.} Let us pack all non-positive helicity fields into $\Psi$ (including the scalar) and all positive helicity fields into $\Phi$. Then the action of Chiral theory in the light-cone gauge reads, schematically
\begin{align}\label{sketch}
    \mathcal{L}&= \Psi_0\square \Psi_0+ \sum_{s>0}\Psi_{-s} \square \Phi_{+s} + c_{+++}\Phi\Phi\Phi+c_{++-}\Phi\Phi\Psi+c_{+--}\Phi\Psi\Psi\,,
\end{align}
where we singled out the kinetic term of the scalar field since it does not have $\Psi\square\Phi$-form. Also, $c_{+\pm\pm}$ keeps track of what kind of a cubic vertex we have in the action. This form can easily be obtained from \eqref{LCaction}. The associated equations of motion are
\begin{align}
    \square \Phi&= c_{++-}\Phi\Phi +c_{+--}\Phi\Psi\,, & \square\Psi&=c_{+++}\Phi\Phi+c_{++-}\Phi\Psi +c_{+--}\Psi\Psi\,.
\end{align}
The vertices in the light-cone gauge have a very simple structure: any three fields can interact as long as the sum of their helicities is positive. Therefore, vertices of type $00+$, $0++$, $+++$ exist for all values of spin (the instanton is associated with helicity $+1$). In particular, we have $0-0-s^+$, $0-1^+-s^+$, $1^+-2^+-s^+$, ... up to $2^+-2^+-s^+$. The only $++-$ vertex where positive-helicity lower spins source a higher spin is $2^+-2^+-3^-$. It is of Yang-Mills type and require colored graviton. The problematic terms, where low-spin fields source higher-spin fields, are now identified as
\begin{align}
    \square \Phi_{+3}&= c_{++-}\Phi_{+2}\Phi_{+2}\,, & \square\Psi_{-s}&=\sum_{s',s''}c_{+++}\Phi_{+s'}\Phi_{+s''}\,.
\end{align}
where $s',s''\in [0,2]$. The light-cone analysis is only preliminary since it does not take into account the higher order vertices required by the covariantization. The source to spin-three vanishes as long as we do not have colored gravitons and we forget about it from now on since colored graviton cannot be a part of the closed low-spin system \cite{Boulanger:2000rq}. In general we observe that low-spin fields can source higher-spin fields and one needs to check whether these sources vanish to claim a consistent low-spin solution.

\paragraph{Back to covariant formulation.} We can compare this structure with the covariant equations of motion. We need to choose some background $\omega_0$. If the spin-two sector is taken to define an empty spacetime, then in Cartesian coordinates we can take $\omega_0$ to consist of 
\begin{align}\label{flatspace}
    e^{AA'}&= \alpha\, dx^{AA'}\,, & \omega^{AA}&=\alpha^2 \,x\fud{A}{C'}dx^{AC'} \,, & \omega^{A'A'}&=0\,,
\end{align}
where $\alpha$ is some parameter with the dimension of inverse length, which was taken $\alpha=1/(2l)$ in Section \ref{sec:fda}. 
Let us assemble $D\equiv d-\omega_0$. The equations \eqref{lowspineqs} give
\besubeqs\label{eq:chiraltheoryA}
\begin{align} 
    D\omega&= \mathcal{V}(\omega, \omega) +\mathcal{V}(\omega_0,\omega,C)+\mathcal{V}(\omega_0,\omega_0,C)\,,\\
    DC&= \mathcal{U}(\omega,C)+ \mathcal{U}(\omega_0,C,C) \,.
\end{align}
\esubeqs
The system above is complete under three assumptions:\footnote{Covariant equations of motion contain vertices of arbitrarily high order, which is, perhaps, the price for covariantization. Therefore, we displayed only the part that is relevant for the (consistent) low-spin solutions.} (i) the fluctuations $\omega$, $C$ contain low spins $s=0,1,2$ only, see \eqref{lowspinan}; (ii) the gravitational sector (both in $\omega$ and in $\omega_0$) is abelian, see \eqref{lowspinan} again; (iii) the low-spin fields do not source higher-spin ones. Indeed, it is easy to see from \eqref{higherV} that on the low-spin solutions already $\mathcal{V}(\omega_0,\omega_0,C,C)=0$ and the higher vertices $\mathcal{V}(\omega_0,\omega_0,C^k)=0$, $k>2$ vanish as well. Likewise, with \eqref{higherU} one observed that $\mathcal{U}(\omega_0,C^k)=0$ for $k>2$. Similarly, $\mathcal{V}(\omega_0,A,C^k)=0$ since $A$ is $y$-independent. Concerning (iii), in general low-spin fields will source higher-spin ones. We will give an argument at the end of Section \eqref{sec:BPSTHiSGRA} that this does not happen for a solution with such a high symmetry as BPST-instanton.

\section{HiSGRA BPST instanton}
\label{sec:BPSTHiSGRA}
After the preliminary work done in the previous Section we restrict ourselves to the truncation of Chiral theory that covers the BPST instanton solution. Therefore, we set the spin-two fluctuations to zero. In Cartesian coordinates the equations simplify to 
\besubeqs
\begin{align}
    dA&= AA -2\alpha^2 d^2x_{B'B'} C^{B'B'}\,,\\
    dC&= [A,C] +2\alpha dx^{AA'}\pl_A \pl_{A'}C+ \mathcal{U}(e,C,C)\,,
\end{align}
\esubeqs
where $[A,C]_{PR}\equiv A\fdu{P}{Q} C_{QR}+A\fdu{R}{Q} C_{PQ}$. It is useful to write down the source in more detail as
\begin{align}\label{mainsource}
\begin{aligned}
        \mathcal{U}(e,C,C)= -2\alpha\, y_B dx^{BB'} e^{q_{23}} \int\Big[ &\plb^2_{B'} C(t_1 y, \bry_2+\bry)  C((1-t_1) y, \bry_3+\bry) \\
    &-
     C((1-t_1) y, \bry_2+\bry)  \plb^3_{B'} C(t_1 y, \bry_3+\bry)\Big]\,,
\end{aligned}
\end{align}
This is the system we need to find a solution of. Note that there is no backreaction onto the spin-two sector. As with the EFT of Section \ref{sec:EFT}, finding the solution can be split into three steps: BPST instanton, scalar field, helicity $-1$ field. We also need to make sure that the source $\mathcal{U}(e,C,C)$ does not activate any higher-spin fields. 

\paragraph{BPST instanton again.} We have already found the FDA form of the BPST instanton. Let us reconsider the problem by taking advantage of the generating functions language. The ansatz for the BPST instanton reads  
\begin{align}
    C_{PP}&= f_3(r^2)\left(
k_{PB^{\prime }}\bar{y}^{B^{\prime }}k_{PC^{\prime }}\bar{y}^{C^{\prime }}\right) \exp{\left[ y^M k_{MM'} \bar{y}^{M'} \right]} = f_3\, z_P z_P\, e^{\sigma}\,,
\end{align}
where we defined $k^{AA'}= w(r^2) r^{AA'}$, $\sigma= y^M k_{MM'} \bar{y}^{M'}$, $z_P= k_{PB'}\bry^{B'}$.  It is straightforward to compute\footnote{Some of the formulas below are given in a raw form to make it easier to track down the origin of various terms.}
\begin{align*}
    dC_{PP}&= z_P z_P f_3' (dx\cdot r) + 2z_P\left( w\, dx_{PM'} \bry^{M'} + \tfrac{w'}{w} z_P (dx\cdot r)\right) f_3 e^\sigma+\\
    &+ z_P z_P  \left(w (y dx\bry) + \tfrac{w'}{w} \sigma (dx\cdot r)\right)f_3 e^\sigma\,,\\
    [A,C]_{PP}&=g' f_3 \left( z_Pz_P (dx \cdot r) -2r^2 w\, z_P dx_{PM'} \bry^{M'} \right)e^\sigma\,,\\
    2e^{AA'} \pl_A\pl_A' C_{PP}&= 2\alpha z_Pz_P\left( (dx\cdot r)(3w+\sigma w) - (y dx \bry) k^2 \right)f_3 e^\sigma+\\
    &-4\alpha k^2  \,z_P dx_{PM'} \bry^{M'} \, f_3 e^\sigma\,,
\end{align*}
where $(y dx \bry)= y_M dx^{MM'}\bry_{M'}$, $(dx\cdot r)=dx^{MM'} r_{MM'}$ and $dk^{AA'}=w' (dx\cdot r) + w dx^{AA'}$. One also needs to use
\begin{align}
    y^M \bry^{M'} k_{MA'}k_{AM'}&= k_{AA'} \sigma- k^2 y_A \bry_{A'}
\end{align}
which is a consequence of $T_{A|B}=T_{B|A}+\epsilon_{AB}T\fdu{C}{C}$ for any spin-tensor $T_{A|B}$. The last ingredient is the source $\mathcal{U}(e,C,C)$, for which it is convenient to rewrite $C$ as
\begin{align}
    C_{PP}&=  f_3(r^2)\, \pl_A^\xi \pl_A^\xi \exp{\left[ (y^M+\xi^M) k_{MM'}  \bar{y}^{M'} \right]}\Big|_{\xi=0}\,,
\end{align}
and we omit $|_{\xi=0}$ in what follows. A simple calculation gives
\begin{align}\label{Scalarsource}
    \tfrac12 \mathcal{U}(e,C,C)_{PR} \epsilon^{PR}&= 2\alpha (ydx \bry) (k^2)^2 ( 3  + \sigma )(f_3)^2 e^\sigma\,,\\
    \mathcal{U}(e,C,C)_{PP}&= \alpha k^2 \Big[ (y_Py_P (dx\cdot r) w - dx\fdu{P}{B'} y^Mk_{MB'} y_P) \tfrac23 k^2(4+\sigma)+\notag\\
    &+\tfrac13(y dx \bry) y_P y_P (k^2)^2-2(y dx \bry)z_P z_P \Big](f_3)^2 e^\sigma\,.
\end{align}
We see that the instanton contributes both to the $su(2)$-singlet ($u(1)$-factor of $u(2)$) and to the $su(2)$ sectors of $u(2)$. It is an important feature of \eqref{mainsource} that it preserves $e^\sigma$, i.e., roughly speaking,
\begin{align}
    \mathcal{U}(dx,\bullet e^\sigma,\bullet e^\sigma)&= \bullet e^\sigma\,,
\end{align}
where $\bullet$ denotes some polynomial prefactors. 
The source reveals three types of structures: (a) the singlet one, which contributes to the scalar field; (b) a source for the helicity $-1$ field that is proportional to $y_\bullet y_\bullet$; (c) a source for the helicity $+1$ field itself. The latter is not an actual source since it does not contribute to the Yang-Mills equations. It accounts for the nonlinearities in \eqref{dCeqACetc} that are due to the Bianchi identities. Indeed, the source for the Yang-Mills equation that is contained in
\begin{align}
  \tfrac12 \bry_{A'}\bry_{A'}&: &  \nabla C^{PP|A'A'}&= 2e_{BB'} C^{PP|B,A'B'B'} + ...
\end{align}
is the coefficient of $\bry \bry$ and there is no such term in $\mathcal{U}(e,C,C)$, the lowest relevant for helicity $+1$ being $\bry\bry (yk\bry)$. Collecting all the relevant terms we have
\begin{align}
    (dx\cdot r) z_Pz_P&: &&\begin{aligned}
        &f_3'+2 f_3 w'/w =g' f+3 +6 \alpha w f_3\,,\\
        &w'=2 \alpha w^2\,,
    \end{aligned}\\
    (ydx \bry) z_Pz_P&: &&f_3 w= -2\alpha k^2 f_3 -2\alpha k^2 b f_3^2\,,\\
    dx_{PM'} \bry^{M'} z_P&: &&2 f_3 w=-2g' f_3 r^2 w-4 \alpha k^2 f_3\,,
\end{align}
of which a unique solution is given by (recall that $\rho \equiv r^2$)
\begin{align}
    w(\xx)&=-\frac{1}{2 \alpha  \left(L^2+\xx\right)}\,, && f_3(\xx)=\frac{L^2}{\xx b}\,.
\end{align}
Here, we have introduced a higher-spin coupling constant $b$ that weights the contribution of the source. This coupling just counts nonlinearities in $C$ and, hence, its effect is easy to track down.

\paragraph{Scalar field.} As the first step towards the EFT of Section \ref{sec:EFT}, we notice that the projection of $\mathcal{U}(e,C,C)$ onto the scalar sector (functions that have equal number of $y$ and $\bry$) is an $su(2)$-singlet. Therefore, we take the following ansatz
\begin{align}
    C_{PR}&= \epsilon_{PR}f(\rho, \sigma)\,, && f=\tilde f(\xx,\sigma) e^\sigma\,. 
\end{align}
A simple calculation gives (note that $[A,C]\equiv0$ here)
\begin{align}
    dC_{PR}&= \epsilon_{PR}\Big[ (dx \cdot r) f'  +\left(\tfrac{w'}{w} (dx\cdot r) \sigma +w (ydx\bry)\right)\pl_\sigma f\Big]\,,\\
    2e^{AA'}\pl_A\pl_{A'}C_{PR}&= 2\alpha\epsilon_{PR}\Big[ w (dx \cdot r) \pl_\sigma f+ (\sigma w (dx\cdot r) -k^2 (y dx\bry))\pl^2_\sigma f\Big]\,.
\end{align}
Adding the scalar source \eqref{Scalarsource} to all of the above we find
\begin{align}
    (dx\cdot r)&: &&f' +\tfrac{w'}{w}\sigma \pl_\sigma f=2\alpha w\pl_\sigma f+2\alpha \sigma w \pl_\sigma^2f\,,
   \\
    (ydx\bry)&: && \pl_\sigma f w=-2\alpha xw^2\pl_\sigma^2 f +2\alpha b x^2 w^4 (3+\sigma)f_3^2 e^\sigma\,.\label{sceq2}
\end{align}
Making ansatz $\tilde f= f_1(\xx)+\sigma h_1(\xx)$ we arrive at
\begin{align}
    (dx\cdot r)&: && \begin{aligned}
        \sigma^2&: && \tfrac{w'}{w} h_1=2\alpha w h_1\,,\\
        \sigma^1&: && h_1'+\tfrac{w'}{w} (f_1+h_1)= 2\alpha w h_1 +2\alpha w (f_1+2h_1)\,,\\
        \sigma^0 &: && f_1'=2\alpha w(f_1+h_1)\,,
    \end{aligned}\\
    (ydx\bry)&: && \begin{aligned}
        \sigma^1&: &&  h_1 w = -2\alpha xw^2h_1 +2\alpha b x^2 w^4 f_3^2\,,\\
        \sigma^0 &: && (f_1+h_1)w =-2\alpha xw^2 (f_1+2h_1) +6\alpha b x^2 w^4f_3^2 \,.
    \end{aligned}
\end{align}
A unique solution is given by 
\begin{align}
    f_1(\xx)&= -\frac{2 L^2+\xx}{b} w^2\,,  && h_1(\xx)=-\frac{L^2}{b} w^2\,.
\end{align}
Note that in the free limit, $b=0$, the right solution is $h_1=f_1=0$. 
The most general solution to \eqref{sceq2} is
\begin{align}
    f&=h_1 \sigma e^\sigma+f_1 e^\sigma+\frac{c_1 \xx e^{\sigma+\frac{L^2 \sigma }{\xx}}}{L^2+\xx}+c_2 \,.
\end{align}
This agrees, of course, with the general solution of the EFT induced by the BPST instanton. In what follows we restrict ourselves to $c_{1,2}=0$. Note that the second exponent is simply $\sigma (1+L^2/\xx)=(2\xx\alpha)^{-1} (y r\bry)$. 

\paragraph{Helicity $\boldsymbol{-1}$ field.} The helicity $-1$ part of the system is the most complicated one because there are two sources: one bilinear in the BPST instanton and another one of type scalar $\times$ instanton. The most general ansatz reads
\begin{align}
    C_{PP}&= y_P y_P f(\xx,\sigma)\,.
\end{align}
As before, we easily find that
\begin{align}
    dC_{PP}&= y_Py_P \Big(f'(dx \cdot r) +\left(\tfrac{w'}{w} \sigma (dx\cdot r) +w(ydx \bry) \right)\pl_\sigma f\Big)\,,\\
    [A,C]&= -g' f \left( 2 T_{PP} - (dx\cdot r) y_Py_P \right)\,,\\
    2e^{AA'}\pl_A\pl_{A'} C&= 4\alpha w(T_{PP})\pl_\sigma f +2\alpha w(dx \cdot r) y_Py_P \pl_\sigma f+\\
    &+2\alpha y_Py_P\left( w (dx \cdot r)\sigma -(ydx \bry) k^2\right) \pl^2_\sigma f\,,
\end{align}
where we introduced $T_{PP}=dx\fdu{P}{M'} y^C r_{CM'} y_P$. 
Next, we need to add the instanton-instanton contribution 
\begin{align*}
    U(e,C_{+1},C_{+1})_{PP}&= \alpha k^2 \Big[ (y_Py_P (dx\cdot r) w - w T_{PP}) \tfrac23 k^2(4+\sigma)+\\
    &\qquad\qquad+\tfrac13(y dx \bry) y_P y_P (k^2)^2 \Big](f_3)^2 e^\sigma\,.
\end{align*}
The last but one, we need to compute the scalar-instanton contribution. In order to do that one can represent the polynomial in $\sigma$ prefactor in the scalar part as
\begin{align}
    f(\sigma)e^\sigma= f(\pl_\tau) e^{\tau \sigma} \Big|_{\tau=1}\,.
\end{align}
The result is 
\begin{align*}
    \mathcal{U}(e,C_0,C_{+1})&+\mathcal{U}(e,C_{+1},C_{0})= -y_Py_P(ydx\bry) \tfrac23 \alpha b (k^2)^2 h_1f_3 e^\sigma+\\
    &+\alpha b\Big( w(dx\cdot r) y_Py_P -w T_{PP}\Big)\left(-\tfrac83  k^2 (f_1+h_1) -\tfrac43  k^2 \sigma h_1 \right)f_3e^\sigma\,.
\end{align*}
Lastly, there is also a plus-minus contribution. It does not source the equations for the physical fields, but is there to account for Bianchi identities. It can be decomposed into three different structures (some details can be found in Appendix \ref{app:negative})
\begin{align}
    \mathcal{U}(e,C_{-1},C_{+1})&+\mathcal{U}(e,C_{+1},C_{-1})=\mathcal{W}_1 (dx \cdot r) y_P y_P+\mathcal{W}_2 (y dx \bry) y_P y_P+\mathcal{W}_3 T_{PP}\,.
\end{align}
Adding up all contributions we find
\begin{align}
    (dx \cdot r) y_P y_P&: && 
    \begin{aligned}
        f' +\tfrac{w'}{w}\sigma \pl_\sigma f&= g'f +2\alpha (w\pl_\sigma f + \sigma w \pl^2_\sigma f) +\\
    &+b\alpha (k^2)^2 \tfrac23 (4+\sigma)w f_3^2e^\sigma+\\
    &+w\alpha b \left(-\tfrac83  k^2 (f_1+h_1) -\tfrac43  k^2 \sigma h_1 \right)f_3e^\sigma    +\mathcal{W}_1   \,,
    \end{aligned}\\
    (y dx \bry) y_P y_P&: && 
    \begin{aligned}
        w \pl_\sigma f&= -2\alpha k^2 \pl_\sigma^2 f -\tfrac23 \alpha b (k^2)^2 h_1 f_3 e^\sigma+\tfrac13 \alpha b (k^2)^3 f_3^2 e^\sigma+\mathcal{W}_2\,,
    \end{aligned}\\
    T_{PP}&: && 
    \begin{aligned}
        0&= -2g'f +4\alpha w \pl_\sigma f +\\
        &-\alpha b w\left(-\tfrac83  k^2 (f_1+h_1) -\tfrac43  k^2 \sigma h_1 \right)f_3e^\sigma+\\
        &-\tfrac23\alpha b w(k^2)^2 (4+\sigma )f_3^2 e^\sigma+\mathcal{W}_3\,.
    \end{aligned}
\end{align}
There is a simpler way to solve this system than just to solve it directly. Indeed, we can proceed along the EFT lines of Section \ref{sec:EFT} and derive the source for the physical field $\Psi^{AB}$. This is done by replacing $dx^{AA'}$ with $\bry^{A'}\pl^A$ and setting $\sigma=0$. Indeed, this operation leads to
\begin{align}
    y_A \bry_{A'} \nabla\fdu{C}{A'}\Psi^{PP|AC}&= ...
\end{align}
The equation for the dynamical field is
\begin{align}
    f'&= g'f +4 \alpha b (k^2)^2 w f_3^2- 4 b \alpha w (k^2)f_3(f_1+h_1)
\end{align}
By comparing its solution to \eqref{genneg} we find that $\kappa_1=-\tfrac{1}{8 \alpha ^2 \sqrt{b}}$ and $\kappa_2=-\tfrac{1}{384 \alpha ^4 b}$. Now, the complete equations of motion are linear in $f$ and only its $f(\xx,\sigma=0)$ component is the actual solution determined by the source. The expansion in $\sigma$ is to express the auxiliary fields
\begin{align}
    C^{PP|A(k+2),A'(k)}
\end{align}
as derivatives of the dynamical one $C_{PP|AA}$. Thanks to the rotation invariance we know that
\begin{align}
    C_{PP}&= \sum_k f_k(\xx) y_P y_P (yr\bry)^k && \label{Cppfk} 
\end{align}
It is easy to see that on replacing $dx^{AA'}$ with $y^A\bry^{A'}$ we annihilate the whole $\mathcal{U}(e,C,C)$ and get an equation that relates neighboring $f_k$ ($\tilde \sigma =yr\bry$)
\begin{align}
    f'(\xx,\tilde\sigma )+g'(\xx) f(\xx,\tilde\sigma )-6 \alpha  \pl_{\tilde\sigma} f(\xx,\tilde\sigma )-2 \alpha  \sigma  \pl_{\tilde \sigma}^2 f(\xx,\tilde \sigma )=0\,.
\end{align}
The first two terms represent $(y\nabla\bry) C$ and the last two $2\alpha y^A \bry^{A'} \pl_A\pl_A'$. An explicit solution can be found in Appendix \ref{app:negative}. It is then possible to check that other equations are satisfied as well. 

The most important information are the values of $\kappa_{1,2}$. Their invariant ratio is $\kappa_1^2/\kappa_2=-6$, which is what higher-spin symmetry does.

\paragraph{No sources for higher spin fields!} Now that the solution to the EFT is obtained we can check if it sources the higher spin fields (including the gravity sector, which we would like to stay frozen to the Minkowski space). The worst possible (in other words, the most interesting) scenario is that the EFT sources higher-spin fields, which can then backreact onto the low spins and so on.

A heuristic argument for why BPST instanton cannot induce any higher spins is the fact that with the data we have it is impossible to write down an ansatz. Indeed, we have $r_{AA'}$ and can afford rotation-invariant functions of type $f(\xx\equiv r^2)$. As a result, we can only write
\besubeqs
\begin{align}
    \phi_{PR}&= \epsilon_{PR}f_1(r^2)\,,\\
    \Psi_{PP|AA}&= \epsilon_{PA}\epsilon_{PA} f_2(r^2)\,,\\
    C_{PP|A'A'}&= r_{PA'}r_{PA'} f_3(r^2)\,,\\
    \Psi_{PP|A(2s)}&= 0\,, \qquad s\neq1\,,\\
    C_{PP|A'(2s)}&=0 \,, \qquad s\neq 1\,.
\end{align}
\esubeqs
One can also check directly that the solution found above, when plugged into $\mathcal{U}(e,C,C)$, does not generate any higher-spin sources. Technically, this is thanks to the fact that all derivatives are of the special $yk\bry$-form, i.e. save for $r_{AA'}$ there are no other vectors involved. If, for example, we have two independent vectors $r_{1,2}^{AA'}$, we could form $r_1^{AA'}r_{2}\fud{A}{A'}$ to be used to construct $C_{A(2s)}$. Therefore, multi-instanton solutions could generate sources for higher-spin fields.

\section{Discussion and Conclusions}
\label{sec:finale}
To sum up first, the BPST instanton turns out to be an exact solution of Chiral theory and its embedding thereinto activates two other fields: the opposite helicity spin-one field and a singlet scalar field. The self-duality condition is not modified by the presence of these two fields. There is a simple EFT that couples these three fields and can be extracted either directly from the equations or by comparing with the known action in the light-cone gauge. What higher-spin symmetry does is to fix the coupling constant in the EFT.\footnote{Let us note that there exists a higher-spin extension of SDYM that is a contraction of Chiral theory \cite{Ponomarev:2017nrr}, which also admits a covariant action \cite{Krasnov:2021nsq}. In this theory the BPST instanton is an exact solution that does not activate any other field. We are grateful to Dmitry Ponomarev for this remark.}

Some obvious extensions of the present work include. (a) deformation of the solutions to (Euclidian) anti-de Sitter space. Indeed, Chiral theory smoothly depends on the cosmological constant. In fact, there are no new couplings that can affect the EFT, the only modification being in that the scalar field acquires the mass such that it is dual to $\Delta=1,2$ operators on the CFT side, depending on the boundary conditions. (b) It looks challenging to embed general ADHM-instantons, i.e. multi-instanton solutions, since they can also activate higher spin fields. (c) It would be interesting to find genuine higher spin instantons, i.e. exact solutions extending the BPST one with higher spin fields.

Another natural question is what is the moduli space of instantons in Chiral theory? It looks plausible that the ADHM construction is a part of it, but it may not cover the genuine higher spin instantons. 

In this regard, there is a simple procedure to get new instanton solutions \cite{Wilczek:1976uy}. One can take $su(N)$ Yang-Mills theory and consider various embeddings of $su(2)$ into $su(N)$, from the fundamental one to the principal one. Each embedding dressed with the basic BPST instanton becomes an instanton as well, however, with a different charge in general. There is little doubt it is still true within $u(N)$-gauged Chiral theory. The option to play with different embeddings of $su(2)$ into $su(N)$ is similar to the one for $3d$ higher spin gravities.  

In the paper we restricted ourselves to the flat space, which, in fact, does not make much difference for the instanton. Within AdS/CFT correspondence Chiral theory is dual to a closed subsector of Chern-Simons matter theories \cite{Sharapov:2022awp} and a natural question is what is the CFT interpretation of the instanton, which is an exact solution of the full dual of Chern-Simons matter theories as well.

Chiral HiSGRA should have a twistor formulation, which is perhaps the best way to formulate self-dual theories and extensions thereof, see \cite{Tran:2021ukl,Tran:2022tft,Herfray:2022prf} for the first steps in this direction. What is the twistor geometric characterization of higher-spin instantons? For the truncations of Chiral theory that lead to higher-spin extensions of SDYM and SDGR analogs of Penrose and Wald theorems were obtained in \cite{Herfray:2022prf}.

We should also make a comment regarding formal solutions of formal HiSGRAs. Here, by formal HiSGRA \cite{Sharapov:2019vyd} we mean FDAs, $d\Phi=Q(\Phi)$, where $Q$ is only constrained by $Q^2=0$ and not by locality. In other words, such $Q$ defines an $L_\infty$-algebra. The $L_\infty$-structure maps begin with some higher-spin algebra's structure constants. Such $L_\infty$-algebras are easy to construct \cite{Sharapov:2019vyd}, in general. The first example dates back to \cite{Vasiliev:1990cm} and several others are known, see  e.g. \cite{Sharapov:2019vyd}. However, a generic $Q$ from the same equivalence class, does not lead to a well-defined field theory, the reason being is that canonical equivalences on the $Q$-manifold/$L_\infty$ side result in non-local field redefinitions from the field theory vantage point. Therefore, only a very limited set of reference frames on the $Q$-manifold side corresponds to well-defined equations. For all formal HiSGRAs this frame is not known,\footnote{There are arguments that this frame may not even exist, see e.g. \cite{Bekaert:2015tva,Maldacena:2015iua,Sleight:2017pcz,Ponomarev:2017qab} and \cite{Neiman:2023orj,Jain:2023juk} for a different view on the problem. It well may be that one needs concepts that go beyond the usual local field theory approach to construct the dual of Chern-Simons matter theories.} except for Chiral HiSGRA, which was also formulated as an FDA. 

Nevertheless, formal HiSGRAs give some nontrivial $L_\infty$-structure, which is stable even under very nonlocal field-redefinitions from the field theory point of view. They can be viewed as ansatze for gauge-invariant interactions with infinitely-many free coefficients left unfixed. One can still look for exact solutions of formal HiSGRAs. In practice, solutions are $Q$-morphisms from simple field theories to a given one.\footnote{For example, one can take a flat connection $\omega_0$ of the spacetime symmetry algebra and map it to $\omega=\omega_0$, $C=0$, which an obvious exact solution.} The first such exact solution was found by Sezgin and Sundell in \cite{Sezgin:2005pv}. Few other solutions were also found, see e.g. \cite{Didenko:2009td,Iazeolla:2011cb,Iazeolla:2017dxc}. However, all solutions are found in the field frame where the interactions are clearly different from actual (holographic) HiSGRAs, see e.g. \cite{Boulanger:2015ova}. The first steps in adjusting the field frame while constructing solutions were taken in \cite{Didenko:2021vdb}. 

Therefore, Chiral HiSGRA provides a unique playground where the solutions can be trusted to all orders. Another useful feature is that Chiral theory does not require nonvanishing cosmological constant and the calculation in the flat space are simpler. In this regard, it should be relatively easy to bring the chiral solutions of \cite{Iazeolla:2007wt} into the local frame.\footnote{Remarkably, \cite{Iazeolla:2007wt} went very close to defining Chiral theory without taking the locality issue \cite{Boulanger:2015ova} into account.} 

Concerning other solutions, it is quite easy to see that Sezgin-Sundell solution \cite{Sezgin:2005pv} has a simple Chiral theory's counterpart in the flat space: a plane wave of the free massless scalar field (including the zero momentum, i.e. just a constant) on the Minkowski background is an exact solution, which is also of low-spin type. Also, the $4d$ BTZ-type solutions discussed in \cite{Aros:2019pgj} do not have any locality problem since the zero-form vanishes. It would also be interesting to generalize the very recent observations in the light-cone gauge \cite{Lipstein:2023pih, Neiman:2023bkq} to Chiral theory. Lastly, since Chiral theory seems to be a natural candidate for celestial holography in view of \cite{Monteiro:2022xwq,Ren:2022sws,Ponomarev:2022atv,Ponomarev:2022ryp,Ponomarev:2022qkx} it would be important to study the asymptotic structure of solutions in flat space.

\section*{Acknowledgments}
\label{sec:Aknowledgements}
We are grateful to Per Sundell for suggesting a different problem, which turned out to be much more difficult and whose transmutation resulted in this project. We are also grateful to Dmitry Ponomarev, Per Sundell and Tung Tran for  useful comments on the draft. This project has received funding from the European Research Council (ERC) under the European Union’s Horizon 2020 research and innovation programme (grant agreement No 101002551).

\begin{appendix}

\section{BPST instanton, technicalities}
\label{app:sdym}
With the help of Chiral theory's vertices given in Section \ref{sec:chiral} one can derive (this is, of course, consistent with \cite{Skvortsov:2022unu})
{\footnotesize\begin{eqnarray*}
\tfrac{1}{k!\left( k+2\right) !}\left[ \left( p_{02}\right) ^{k}\left(
q_{02}\right) ^{k+2}\omega C-\left( p_{01}\right) ^{k}\left( q_{01}\right)
^{k+2}C\omega \right] &=&\frac{1}{k!\left( k+2\right) !}\left[
A,l^{k+2}C_{A(k),A^{\prime }(k+2)}\right] \left( y^{A}\right) ^{k}\left( 
\bar{y}^{A^{\prime }}\right) ^{k+2}\ , \\
\tfrac{1}{k!\left( k+2\right) !}\left[ p_{12}q_{12}\left( p_{02}\right)
^{k}\left( q_{02}\right) ^{k+2}\omega C+p_{12}q_{12}\left( p_{01}\right)
^{k}\left( q_{01}\right) ^{k+2}C\omega \right] &=&\tfrac{1}{k!\left(
k+2\right) !}l^{-1}e^{BB^{\prime }}l^{k+3}C_{A(k)B,A^{\prime }(k+2)B^{\prime
}}\left( y^{A}\right) ^{k}\left( \bar{y}^{A^{\prime }}\right) ^{k+2}\ .
\end{eqnarray*}}%
Here we also replaced $e^{AA'}$ with $\tfrac{1}{2l} e^{AA'}$ to consistently introduce dimensionful parameter $l$. Likewise, for the cubic vertices we have 
{\footnotesize\begin{eqnarray*}
&&\tfrac{2}{\left( n+1\right) !\left( k-n-1\right) !\left( k+1\right) !\left(
k-n+1\right) }\left[ p_{01}q_{12}\left( p_{02}\right) ^{n}\left(
q_{02}\right) ^{n+1}\left( p_{03}\right) ^{k-n-1}\left( q_{03}\right)
^{k-n+1}\right.  \notag \\
&&\ \ \ \ \ \ \ \ \ \ \ \ \ \ \ \ \ \ \ \ \ \ \ \ \ \ \ \ \ \ \ \ \ \ \ \ \
\ \ \ \ \ \ \ \ \ \ \ \left. -p_{01}q_{13}\left( p_{03}\right) ^{n}\left(
q_{03}\right) ^{n+1}\left( p_{02}\right) ^{k-n-1}\left( q_{02}\right)
^{k-n+1}\right] \omega CC  \notag \\
&=&\tfrac{1}{\left( n+1\right) !\left( k-n-1\right) !\left( k+1\right)
!\left( k-n+1\right) }l^{-1}e_{A}{}^{B^{\prime }}\left[ l^{n+2}C_{A(n),A^{%
\prime }(n+1)B^{\prime }},l^{k-n+1}C_{A(k-n-1),A^{\prime }(k-n+1)}\right] \
\left( y^{A}\right) ^{k}\left( \bar{y}^{A^{\prime }}\right) ^{k+2}\ .
\end{eqnarray*}}%
We substitute:
\begin{equation*}
A_{AA|BB^{\prime }}=-\varepsilon _{AB}r_{AB^{\prime }}g^{\prime }\ ,
\end{equation*}
\begin{equation*}
C_{CC|A(k),A^{\prime }(k+2)}=\left( r_{CA^{\prime }}\right) ^{2}\left(
r_{AA^{\prime }}\right) ^{k}f_{k}\text{\ \ i.e.\ \ }C_{AA|A(k),A^{\prime
}(k+2)}=\left( r_{AA^{\prime }}\right) ^{k+2}f_{k}\ ,
\end{equation*}
where $g$ and $f_k$ are functions of $\rho\equiv r^2$, and $g$ is already obtained in Section \ref{sec:BPST}.\footnote{We slightly abuse the notation here that $f_k$ in this appendix is different from elsewhere e.g.\ (\ref{Cppfk}).}
Then we get the following system of equations by projecting the equations of motion onto various irreducible components
\begin{eqnarray*}
f_{k}^{\prime } &=&g^{\prime }f_{k}+f_{k+1}\ , \\
0 &=&-\frac{k+3}{k+2}g^{\prime }f_{k}+2\left( k+3\right) \left( k-1\right)
!\sum_{n=0}^{k-1}\frac{\left( k-n\right) }{\left( k-n+1\right) !\left(
n+2\right) !}f_{n}f_{k-n-1}\ , \\
-\frac{k+3}{k+2}\left[ \left( k+3\right) \frac{1}{r^{2}}f_{k}+f_{k}^{\prime }%
\right]  &=&\frac{k+3}{k+2}g^{\prime }f_{k}+\frac{2\left( k+3\right) \left(
k+1\right) !}{k}\sum_{n=0}^{k-1}\frac{\left( k-n\right) }{\left(
k-n+1\right) !\left( n+2\right) !}f_{n}f_{k-n-1}\ , \\
-\frac{k+3}{k+2}\left[ \left( k+3\right) \frac{1}{r^{2}}f_{k}+f_{k}^{\prime }%
\right]  &=&2\frac{k+3}{k+2}g^{\prime }f_{k}+2\left( k+3\right)
k!\sum_{n=0}^{k-1}\frac{\left( k-n\right) }{\left( k-n+1\right) !\left(
n+2\right) !}f_{n}f_{k-n-1}\ .
\end{eqnarray*}%
It is obvious that the last three equations are not independent. The above
four equations can be simplified as (it is assumed $k\geq 1$)%
\begin{eqnarray}
h_{k+1} &=&\frac{1}{k+3}\left( h_{k}^{\prime }-g^{\prime }h_{k}\right) \ ,
\label{gheq1} \\
-g^{\prime } &=&\frac{1}{k+2}\frac{h_{k}^{\prime }}{h_{k}}+\frac{k+3}{k+2}%
\frac{1}{r^{2}}\ ,  \label{gheq2} \\
\left( k+3\right) \frac{1}{r^{2}}h_{k}+h_{k}^{\prime } &=&-\frac{k+2}{k}%
\sum_{n=0}^{k-1}h_{n}h_{k-n-1}\ ,  \label{gheq3}
\end{eqnarray}
where $h_k\equiv \frac{1}{(k+2)!} f_k$. As it can be checked, the solution is given by
\begin{equation}
h_{k}=\left( -1\right) ^{k+1}\frac{L^{2}}{r^{2}\left( L^{2}+r^{2}\right)
^{k+2}}\ ,\text{ \ for \ }k=0,1,2,\cdots \,,  \label{hkinr}
\end{equation}%
which leads to the generating function (\ref{Ckarb}). 

\section{Negative helicity, auxiliary fields}
\label{app:negative}
In this Section we collect some technicalities that are needed to solve the complete FDA equations for the helicity $-1$ field. This is also a good illustration of how simple it is to just solve for the dynamical fields, see Section \ref{sec:EFT} as compared to solving for the full package of auxiliary fields. First, we present the generating functions that correspond to all four parts of the solution \eqref{genneg}. Which one is which can be seen from the dependence on $c_1$, $c_2$ and $\kappa_{1,2}$. 
\begin{align*}
    \frac{f}{\alpha c_1}&= \frac{12  L^2 w}{\xx}\int_{\Delta_2} \exp\left[-\frac{\sigma  t_1}{2 \alpha  w \xx}\right]\Big[ { \left(L^2-\xx\right)}-\delta(t_2-t_1){   L^2 } \Big]-\tfrac{2   L^6 w }{\xx^2}e^{-\frac{\sigma }{2 \alpha  w \xx}}-2    w \xx\,.
\end{align*}
Here, we recall that $\Delta_2$ is the two-dimensional simplex, $0\leq t_1\leq t_2\leq1$. Other three solutions read
\begin{align*}
    f&= -\alpha  \kappa _1 c_s\frac{8 L^2 w }{3 \xx^2} \exp \left[{-\frac{\sigma }{2 \alpha  w \xx}}\right]\,,
\end{align*}
\begin{align*}
    \frac{f}{\kappa_1^2}&= \tfrac{128}{5} \alpha ^3 L^2 w^3\int_{\Delta_2} e^{\sigma  t_1-\frac{\sigma  \left(1-t_1\right)}{2 \alpha  w \xx}}\Big[-\frac{L^2-\xx}{\xx}-2\delta(t_2-t_1)\Big]+\tfrac{64 \alpha  L^2 w }{15 \xx^2}e^{-\frac{\sigma }{2 \alpha  w \xx}}-\tfrac{128}{3} \alpha ^4 L^2 e^{\sigma } w^4 \xx\,,
\end{align*}
\begin{align*}
    \frac{f}{\kappa_2}&= \tfrac{128}{5} \alpha ^3 L^2 w^3\int_{\Delta_2} e^{\sigma  t_1-\frac{\sigma  \left(1-t_1\right)}{2 \alpha  w \xx}}\Big[\frac{3 (L^2 - \xx)}{2 \xx}+3\delta(t_2-t_1)\Big]+\tfrac{8 \alpha  L^2 w }{5 \xx^2}e^{-\frac{\sigma }{2 \alpha  w \xx}}+128 \alpha ^4 L^2 e^{\sigma } w^4 \xx\,.
\end{align*}
In order to compute the contribution from $\mathcal{U}(e,C,C)$ of type plus-minus, we can utilize the same trick 
\begin{align}
    f(\sigma)e^\sigma= f(\pl_\tau) e^{\tau \sigma} \Big|_{\tau=1}\,.
\end{align}
This way, we get the following projections onto the three tensor structures
\begin{align*}
    \mathcal{W}_1&=4 \alpha  \sigma  w \left(\left(t_1-1\right){}^2 \left(\sigma  t_1+2\right) e^{\sigma  \left(\tau -(\tau -1) t_1\right)}-\sigma  \tau  t_1^3 e^{\sigma +\sigma  (\tau -1) t_1}\right)\,,\\
    \mathcal{W}_2&=4 \alpha  w^2 \xx \left(\sigma  \tau  t_1^3 e^{\sigma +\sigma  (\tau -1) t_1}-\left(t_1-1\right){}^2 \left(\sigma  t_1+1\right) e^{\sigma  \left(\tau -(\tau -1) t_1\right)}\right)\,,\\
    \mathcal{W}_3&=-\mathcal{W}_1\,,
\end{align*}
This still needs to be multiplied by $f_3$ and the integral over $\Delta_2$ is implied. It is a straightforward but lengthy calculation to check that each of the four solutions above solves the equations for the negative helicity field.

\end{appendix}

\footnotesize
\providecommand{\href}[2]{#2}\begingroup\raggedright\endgroup

\end{document}